\documentclass[pdflatex,sn-aps,iicol]{sn-jnl}% Basic Springer Nature Reference Style/Chemistry Reference Style
\usepackage{graphicx}%
\usepackage{multirow}%
\usepackage{amsmath,amssymb,amsfonts}%
\usepackage{amsthm}%
\usepackage{mathrsfs}%
\usepackage[title]{appendix}%
\usepackage[table]{xcolor}%
\usepackage{textcomp}%
\usepackage{manyfoot}%
\usepackage{booktabs}%
\usepackage{algorithm}%
\usepackage{algorithmicx}%
\usepackage{algpseudocode}%
\usepackage{listings}%
\RequirePackage{tkz-euclide}
\usetikzlibrary{calc, shapes, backgrounds, bending, decorations.pathmorphing, positioning, decorations.pathreplacing, spy}
\usepackage{pgfplots}
\usepackage{tikz-layers}
\usepgflibrary{arrows}

\pgfplotsset{compat=1.18}

\usepackage{tabularx}
\usepackage{arydshln}

\usepackage{array}%
\usepackage{float}%
\usepackage{makecell}%

\theoremstyle{thmstyleone}%
\theoremstyle{thmstyletwo}%

\theoremstyle{thmstylethree}%
\newcommand{\wrap}[1]{#1}

\begin{document}

\title[Article Title]{Numerical and Experimental Evaluation of Chip Evacuation and Lubricant Flow using Optimized Drill Heads for Ejector Deep Hole Drilling}

\author[1]{\fnm{Nuwan} \sur{Rupasinghe}}
\author[2]{\fnm{Sebastian} \sur{Michel}}
\author[1]{\fnm{Andreas} \sur{Baumann}}
\author[2]{\fnm{Julian} \sur{Gerken}}
\author[2]{\fnm{Samuel} \sur{Gülde}}
\author[2]{\fnm{Dirk} \sur{Biermann}}
\author*[1]{\fnm{Peter} \sur{Eberhard}}\email{peter.eberhard@itm.uni-stuttgart.de}

\affil*[1]{\orgdiv{Institute of Engineering and Computational Mechanics}, \orgname{University of Stuttgart}, \orgaddress{\street{Pfaffenwaldring 9}, \postcode{70569} \city{Stuttgart}, \country{Germany}}}

\affil[2]{\orgdiv{Institute of Machining Technology}, \orgname{University of Dortmund}, \orgaddress{\street{Baroper Street 303}, \postcode{44227} \city{Dortmund}, \country{Germany}}}


\abstract {Ejector deep hole drilling offers great potential to utilize the typical advantages of deep hole drilling processes on conventional machining centers in a cost-effective and resource-efficient manner. However, maintaining reliable chip evacuation and stable process conditions often relies on high flow volumes of metalworking fluid, resulting in considerable energy consumption in industrial settings. Therefore, to analyze the highly sophisticated chip evacuation dynamics of the process, two flow-optimized drill heads and a reference drill head were evaluated with smoothed particle hydrodynamics simulation using experimentally obtained chip shapes. In addition, modified drill heads were additively manufactured and experimentally investigated to validate the numerical results and to determine the positive effect on the necessary fluid flow for a stable ejector drilling process. The modifications aim to improve chip evacuation by reducing vortex formation and optimizing flow conditions near the cutting zone. Therefore, the minimum volume flow required for a stable drilling process without chip clogging is reduced, leading to an energy-efficient sustainable ejector drilling process.}

\keywords{ejector deep hole drilling, chip evacuation, tool modification, smoothed particle hydrodynamics (SPH), cutting tool}

\maketitle

\section{Introduction}\label{sec1}

Deep hole drilling (DHD) processes are characterized by their high length-to-diameter ratios of $\mathrm{L/D \geq 10}$ and high achievable bore hole qualities. The application fields are, e.g., automotive, aerospace, or medical industries, where they are among others used to manufacture hydraulic cylinders, gear shafts or valve guides \cite{Biermann2018}. The classical DHD methods single-lip drilling, drilling with a single-tube (BTA drilling), and double-tube system (ejector drilling) are characterized by the high level of productivity that is achieved due to high feed rates and a continuous drilling process. In consequence, the chip and heat removal from the cutting zone is essential. This is typically achieved by a metalworking fluid (MWF) flow under high volume flow rates or high pressures. In contrast to single-lip DHD, ejector drilling removes the chips through the inside of the tool. It has a cooling MWF supply using the Venturi-based ejector effect with a supply at much lower pressures and without the need for a seal between the face of the workpiece and the drilling tool like in BTA deep hole drilling. Therefore, with ejector deep hole drilling (EDHD), conventional machining centers are able to utilize the advantages of DHD such as a high material removal rate with high bore hole qualities in an efficient manner and without special equipment \cite{VDI2006}.

\begin{figure}[h]
  \centering
  \def\svgwidth{\columnwidth}  % or any width you want
  
\begingroup%
  \makeatletter%
  \providecommand\color[2][]{%
    \errmessage{(Inkscape) Color is used for the text in Inkscape, but the package 'color.sty' is not loaded}%
    \renewcommand\color[2][]{}%
  }%
  \providecommand\transparent[1]{%
    \errmessage{(Inkscape) Transparency is used (non-zero) for the text in Inkscape, but the package 'transparent.sty' is not loaded}%
    \renewcommand\transparent[1]{}%
  }%
  \providecommand\rotatebox[2]{#2}%
  
    % Set font size
  \footnotesize

  \newcommand*\fsize{\dimexpr\f@size pt\relax}%
  \newcommand*\lineheight[1]{\fontsize{\fsize}{#1\fsize}\selectfont}%
  \ifx\svgwidth\undefined%
    \setlength{\unitlength}{255.12000275bp}%
    \ifx\svgscale\undefined%
      \relax%
    \else%
      \setlength{\unitlength}{\unitlength * \real{\svgscale}}%
    \fi%
  \else%
    \setlength{\unitlength}{\svgwidth}%
  \fi%
  \global\let\svgwidth\undefined%
  \global\let\svgscale\undefined%
  \makeatother%
  \begin{picture}(1,0.49391465)%
    \lineheight{1}%
    \setlength\tabcolsep{0pt}%
    \put(0,0){\includegraphics[width=\unitlength,page=1]{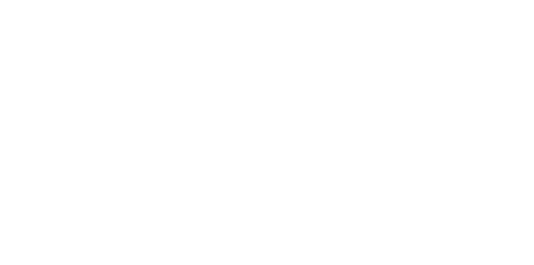}}%
    \put(0.01354657,0.45138565){\makebox(0,0)[lt]{\lineheight{1.25}\smash{\begin{tabular}[t]{l}1. first guide pad\end{tabular}}}}%
    \put(0.01354657,0.407179){\makebox(0,0)[lt]{\lineheight{1.25}\smash{\begin{tabular}[t]{l}4. inner cutting edge\end{tabular}}}}%
    \put(0.37,0.45138565){\makebox(0,0)[lt]{\lineheight{1.25}\smash{\begin{tabular}[t]{l}2. second guide pad\end{tabular}}}}%
    \put(0.37,0.40676492){\makebox(0,0)[lt]{\lineheight{1.25}\smash{\begin{tabular}[t]{l}5. outer cutting edge\end{tabular}}}}%
    \put(0.73,0.45138565){\makebox(0,0)[lt]{\lineheight{1.25}\smash{\begin{tabular}[t]{l}3. third guide pad\end{tabular}}}}%
    \put(0.73,0.407179){\makebox(0,0)[lt]{\lineheight{1.25}\smash{\begin{tabular}[t]{l}6. chip mouth\end{tabular}}}}%
    \put(0.01354657,0.3576964){\makebox(0,0)[lt]{\lineheight{1.25}\smash{\begin{tabular}[t]{l}7. MWF outlet bores\end{tabular}}}}%
    \put(0,0){\includegraphics[width=\unitlength,page=2]{Fig1.pdf}}%
    \put(0.03142051,0.25233422){\makebox(0,0)[lt]{\lineheight{1.25}\smash{\begin{tabular}[t]{l}2\end{tabular}}}}%
    \put(0,0){\includegraphics[width=\unitlength,page=3]{Fig1.pdf}}%
    \put(0.12662276,0.31696263){\makebox(0,0)[lt]{\lineheight{1.25}\smash{\begin{tabular}[t]{l}3\end{tabular}}}}%
    \put(0,0){\includegraphics[width=\unitlength,page=4]{Fig1.pdf}}%
    \put(0.56922233,0.25186385){\makebox(0,0)[lt]{\lineheight{1.25}\smash{\begin{tabular}[t]{l}2\end{tabular}}}}%
    \put(0,0){\includegraphics[width=\unitlength,page=5]{Fig1.pdf}}%
    \put(0.56922233,0.31696263){\makebox(0,0)[lt]{\lineheight{1.25}\smash{\begin{tabular}[t]{l}4\end{tabular}}}}%
    \put(0,0){\includegraphics[width=\unitlength,page=6]{Fig1.pdf}}%
    \put(0.95182657,0.31696263){\makebox(0,0)[lt]{\lineheight{1.25}\smash{\begin{tabular}[t]{l}3\end{tabular}}}}%
    \put(0,0){\includegraphics[width=\unitlength,page=7]{Fig1.pdf}}%
    \put(0.95182657,0.15656752){\makebox(0,0)[lt]{\lineheight{1.25}\smash{\begin{tabular}[t]{l}5\end{tabular}}}}%
    \put(0,0){\includegraphics[width=\unitlength,page=8]{Fig1.pdf}}%
    \put(0.95182657,0.07923921){\makebox(0,0)[lt]{\lineheight{1.25}\smash{\begin{tabular}[t]{l}1\end{tabular}}}}%
    \put(0,0){\includegraphics[width=\unitlength,page=9]{Fig1.pdf}}%
    \put(0.56922233,0.17829847){\makebox(0,0)[lt]{\lineheight{1.25}\smash{\begin{tabular}[t]{l}7\end{tabular}}}}%
    \put(0,0){\includegraphics[width=\unitlength,page=10]{Fig1.pdf}}%
    \put(0.23207902,0.07923921){\makebox(0,0)[lt]{\lineheight{1.25}\smash{\begin{tabular}[t]{l}5\end{tabular}}}}%
    \put(0,0){\includegraphics[width=\unitlength,page=11]{Fig1.pdf}}%
    \put(0.03142051,0.16578672){\makebox(0,0)[lt]{\lineheight{1.25}\smash{\begin{tabular}[t]{l}4\end{tabular}}}}%
    \put(0,0){\includegraphics[width=\unitlength,page=12]{Fig1.pdf}}%
    \put(0.03142051,0.07923921){\makebox(0,0)[lt]{\lineheight{1.25}\smash{\begin{tabular}[t]{l}1\end{tabular}}}}%
    \put(0,0){\includegraphics[width=\unitlength,page=13]{Fig1.pdf}}%
    \put(0.32229539,0.07923921){\makebox(0,0)[lt]{\lineheight{1.25}\smash{\begin{tabular}[t]{l}6\end{tabular}}}}%
    \put(0,0){\includegraphics[width=\unitlength,page=14]{Fig1.pdf}}%
    \put(0.95182657,0.24349132){\makebox(0,0)[lt]{\lineheight{1.25}\smash{\begin{tabular}[t]{l}6\end{tabular}}}}%
    \put(0,0){\includegraphics[width=\unitlength,page=15]{Fig1.pdf}}%
    \put(0.56922233,0.10605012){\makebox(0,0)[lt]{\lineheight{1.25}\smash{\begin{tabular}[t]{l}6\end{tabular}}}}%
    \put(0,0){\includegraphics[width=\unitlength,page=16]{Fig1.pdf}}%
  \end{picture}%
\endgroup%

  \caption{Ejector drill head Botek type 62 (D = 30mm)}
  \label{fig:drill_head}

\end{figure}

Illustrated in Figure~\ref{fig:drill_head} is an ejector drill head with indexable cutting inserts and guide pads. Here, the cutting edge is divided into two parts as outer cutting edge and inner cutting edge for a drill diameter of $\textrm{D = 30\,mm}$. With the inner cutting edge shifted by $\textrm{180}^{\circ}$ and positioned directly in the center on the opposite side to the outer cutting edge, the two are aligned to provide minimal overlap. The asymmetrical arrangement of the cutting edges results in strong radial forces during the drilling process, which are transferred to the bore hole wall by the guide pads. This configuration provides a self-guiding effect thus improving the straightness and roundness of the bore hole. However, the contact between the bore hole wall and the guide pads generates significant thermal and mechanical loads, making effective cooling and lubrication essential for process stability, tool life, and bore hole quality. Hence, the MWF is supplied via the annular channel between the double tube system and emerges through the outlet bores at the drill head. The MWF circulates around the cutting zone and the resulting chip-MWF mixture is flushed away and evacuated through the chip mouths in the drill head and through the inner tube. This process is further accelerated and supported by the lower pressure inside the inner tube resulting from the ejector effect. This ensures that chips have no contact with the bore hole wall and, therefore, prevents the chips from scratching or damaging the bore hole surface \cite{Biermann2018}.

Since the operating pressure of the MWF at the inlet is lower compared to other DHD methods~\cite{Biermann2018}, efficient removal of chips from the cutting zone is crucial to avoid chip clogging, which otherwise leads to poor bore hole quality and significant tool wear. The ejector effect and the reliable chip removal from the bore hole are strongly dependent on the MWF flow and the design of the tool \cite{Astakhov1995}. To ensure high process stability and avoid chip clogging, high volume flows are currently used in industrial applications, leading to a high energy consumption of the process \cite{Astakhov2014}.
Optimizing the drill heads with geometric modifications and a better understanding of the process limits hold a great potential to reduce the necessary volume flow and improve the efficiency of the ejector deep hole drilling process. 

Therefore, numerical simulations using the smoothed particle hydrodynamics (SPH) method were successfully performed to investigate the design modifications of the outlet bores to improve the MWF supply to the contact zone~\cite{Gerken2023}. However, in previous experimental and numerical investigations, it was possible to observe a vortex formation near the outer cutting edge \cite{Fritsching2023}. This vortex can affect chip evacuation as chips are caught in the vortex, leading to delayed chip removal \cite{Baumann2024}. A chip removal slower than their creation can cause chip blockage and results in drill breakage due to an increased torque \cite{Kumar2021}. To mitigate the vortex formation at the outer cutting edge, different design variants of the ejector drill head and their influence on fluid flow have been numerically investigated \cite{Baumann2024}. 

In the present work, the chip evacuation of the two most promising modifications from these previous results are analyzed in detail using SPH simulations. The modified drill head designs are manufactured using an additive manufacturing process and machining to prepare the functional surfaces for the cutting edges and the guide pads. Experimental investigations are carried out to validate the simulations and to determine the positive effect of the modifications on the chip removal and the necessary MWF volume flow rate for a stable ejector DHD process.

\maketitle
\section{Simulation Model and Setup}\label{sec2}

In order to accurately model the physical behavior and characteristics of chip movement and coolant flow in the ejector drill head, the SPH method is used. In the following, first the SPH method is described. Afterwards, the simulation model settings, the applied boundary conditions, and the investigated drill head designs are explained.

\subsection{Smoothed Particle Hydrodynamics}\label{subsec1}

SPH is a meshless Lagrangian particle method well-suited for problems involving complex geometries, moving boundaries, and multiphase interactions such as the high-pressure coolant and chip dynamics found in EDHD. Unlike traditional Eulerian grid-based methods, SPH discretizes the fluid domain into a set of interpolation points called "particles" that move with the flow, making it particularly effective for modeling transient, free-surface, and evolving fluid structures. Due to its meshless nature, it can be efficiently coupled with the Discrete Element Method (DEM) for solving the complex chip fluid interaction inside the drill head.

In its standard SPH formulation, a property of a particle in the fluid domain is discretized by weighted summation of a finite
set of neighboring particles, providing the particle approximation \cite{Liu2010}

\begin{equation}
\label{eqn:Chap2SPHapproxEqn}
    A(\mathbf{r}) = \sum_{b}\frac{m_{b}}{\rho_{b}}A_{b}W(\mathbf{r}-\mathbf{r}_b,h),
\end{equation}
where $A(\mathbf{r})$ is the particle property which needs to be approximated, $\mathbf{r}$ is a position vector, and each particle carries a mass $m$ and density $\rho$. The summation using the kernel function $W(\mathbf{r}-\mathbf{r}_{b},h)$ depends on the support domain defined by the smoothing length $h$, localizing the problem domain to effective neighboring particles $b$.

In this study, the weakly compressible SPH (WCSPH) formulation is applied to solve the Navier-Stokes equations. In their Lagrangian form the continuity equation is

\begin{equation}
\label{eqn:Chap2ContinuityEqn}
    \frac{d \rho}{dt} = - \rho\nabla \cdot \mathbf{v},
\end{equation} 

describing the conservation of the mass. Here, $\mathrm{t}$ is the time, and $\mathbf{v}$ is the velocity. The momentum in the fluid is conserved by

\begin{equation}
\label{eqn:Chap2MomentumEqn}
   \frac{d\mathbf{v}}{dt} = \frac{1}{\rho} ( - \nabla p + \mu \nabla ^2 \mathbf{v} + \mathbf{f} + \nabla R),
\end{equation}

which are the Navier-Stokes (N-S) equations. Here, $p$ is pressure, and $\mu$ is the dynamic viscosity~\cite{McLean2012}. The external forces, and the Reynolds turbulent stresses are represented by $\mathbf{f}$  and $\mathbf{R}$. Further, the relation between position $\mathbf{r}$ and velocity $\mathbf{v}$ is provided by the kinematics equation,

\begin{equation}
\label{eqn:Chap2KinematicEqn}
    \frac{d\mathbf{r}}{dt} = \mathbf{v}.
\end{equation}

As indicated in Eq. \ref{eqn:Chap2ContinuityEqn}, the fluid is not strictly incompressible, permitting small density fluctuations. Tait’s equation of state \cite{{Murnaghan1944}} is applied to enforce quasi-incompressibility by introducing a restoring force that resists the concentration of the fluid, expressed by

\begin{equation}
\label{eqn:Chap1EquationOfState}
    p = \frac{c_{o}^2\rho_{o}}{\gamma}\left [ \left (\frac{\rho}{\rho_{o}}\right )^{\gamma} - 1 \right ].
\end{equation}

Here, $\rho_{o}$ is the reference density, and the polytropic index denoted by $\gamma$ is set to 7 for water \cite{Monaghan1994}. The value of the numerical speed of sound $c_{o}$ is chosen as 10 times the maximum expected velocity of the fluid flow which in turn restricts the density fluctuations to less than $1\%$ \cite{Violeau2016}.

Therefore, Equations~\ref{eqn:Chap2ContinuityEqn} and \ref{eqn:Chap2MomentumEqn} are solved at discrete particle positions and the discretized SPH formulation of the  N-S equation is given by
\begin{eqnarray}
    \label{eqn:Chap2SPHNavStokesEqn}
    \left. \frac{d\mathbf{v}}{dt} \right|_{a} = &&\mathbf{g} - \sum_{b}m_{b}\left[ \left (\frac{p_{a}}{\rho_{a}^{2}}+ \frac{p_{b}}{\rho_{b}^{2}} \right ) \right . \nonumber \\
    &&\left .-\frac{1}{\rho_{a}\rho_{b}}\frac{(\mu_{a}+\mu_{b}) \mathbf{v}_{ab} \cdot \mathbf{r}_{ab}}{||\mathbf{r}_{ab}||^{2} + 0.01h^{2}} \right] \nabla_a W_{ab},
\end{eqnarray}
and the continuity equation reads
\begin{equation}
    \label{eqn:Chap2SPHContinuityEqn}
    \left. \frac{d\rho}{dt} \right|_{a} = \rho_{a}\sum_{b}\frac{m_{b}}{\rho_{b}}\mathbf{v}_{ab} \cdot \nabla_a W_{ab},
\end{equation}
where $\mathbf{v}_{ab}$ is the relative velocity and $W_{ab}$ denotes the kernel function $W(\mathbf{r}_{a} - \mathbf{r}_{b},h)$ in its simplified notation and $\nabla_a$ describes the gradient calculated with respect to a particle $a$ and its position. In addition, $\mathbf{r}_{ab}$ is the distance between particles, and $\mu_{a}$ and $\mu_{b}$ is the dynamic viscosity of the respective particles $a$ and $b$. 

SPH approximations become problematic near open free surfaces due to incomplete kernel support inherent to the kernel function's implementation. Therefore, the kernel gradient is replaced by a modified version to improve the numerical accuracy \cite{Bonet1999}. Similarly, the continuity equation~\ref{eqn:Chap2SPHContinuityEqn} is modified by adding an artificial diffusion term to remove the spurious numerical oscillations in the pressure field, which is one of the drawbacks of the standard WCSPH \cite{Molteni2009}. Additionally, an artificial viscous term is added to improve the numerical stability of the simulation as described by~\cite{Monaghan1994}.

The SPH fluid model is coupled with DEM to capture the rigid chip dynamics during the drilling process. The solid boundaries of the chips, drill tool, and bore hole are defined using triangular meshes and the repulsive forces from the interacting particles are acting normal to the triangles. Hence, these interactions between the SPH fluid particles and the solid structures are modeled using the Modified Lennard-Jones Potential~\cite{Mueller2004}. This repulsive force formulation ensures the impermeability of the solid boundaries, while remaining computationally efficient~\cite{Kazaz2016}. The boundary interaction force $F_{d}$ between a fluid particle and a solid boundary at distance $d$ is defined as

\begin{equation}
\label{eqn:Chap2ModLenJonPotential}
    F_{d} = \begin{cases}\psi\frac{(R-d)^{4} - (R-s)^{2}(R-d)^{2}}{R^{2} s (2R-s)}
    & \text{if~} d \leq R, \\ 
    0 & \text{otherwise.}
\end{cases}
\end{equation}
Here, $R$ is the maximum range for interactions and the parameter $s$ denotes the distance, where the force $F_{d}$ switches from repulsive to attractive. The user defined stiffness parameter $\psi$ controls the force magnitude at contact and prevents the singularities as the distance disappears allowing purely repulsive interactions. The movement of the chips, which are modeled as DEM particles, are governed by the Newton-Euler equations~\cite{Schielen2014} represented by
\begin{equation}
\label{eqn:Chap2NewtonEulerTranslation}
\mathbf{F}_{i} = m_{i} \mathbf{a}_{i}\,,
\end{equation}
\begin{equation}
\label{eqn:Chap2NewtonEulerRotational}
\mathbf{T} = \mathbf{I} \dot{\boldsymbol{\omega}} + \boldsymbol{\omega} \times (\mathbf{I} \boldsymbol{\omega})\,,
\end{equation}\\
with the force represented by $\mathbf{F}_{i}$, mass of single solid particle $m_{i}$, acceleration is $\mathbf{a}_{i}$, the inertia tensor is $\mathbf{I}$, and the torques as $\mathbf{T}$. Contacts between DEM particles are handled using a unilateral penalty method, employing a linear spring-damper model to mitigate particle overlap during collisions.

An explicit second order predictor-corrector leapfrog integrator is applied for the time stepping \cite{Monaghan2003}. The time step size $\Delta t$ is controlled by the Courant-Friedrichs-Lewy (CFL) conditions, restricting the time step, so that the propagation of the information remains within the limits of the spatial discretization. Here, $\alpha_c$ denotes the application dependent Courant number, thus
\begin{equation}
\label{eqn:Chap2CFL}
\Delta t \leq \Delta t_{\text{CFL}} = \alpha_c \frac{h}{c_{o}}\,.
\end{equation}

\subsection{Simulation Setup}\label{subsec2}
\begin{figure*}[htb]
    \centering

    \noindent
\def\wimage{.2\textwidth}
\def\himage{.2\textwidth}
\def\wframe{0cm}
\def\hframe{0cm}

\begin{tikzpicture}
  %\footnotesize

  % \draw (0,0) node [draw=black] {My text 4};

  % Center Node
  \node[draw=none, align=center] (center_left) at (-.8*\wimage-.55cm,0) {};
  \node[draw=none, align=center] (center) at (0-.5cm,0) {};
  \node[draw=none, align=center] (center_right) at (.8*\wimage-.45cm,0) {};

  % Images
  \node[anchor=south east, xshift=-\wframe, yshift = \hframe] at (center_left) {\includegraphics[trim=115 45 375 100, clip, height=\himage]{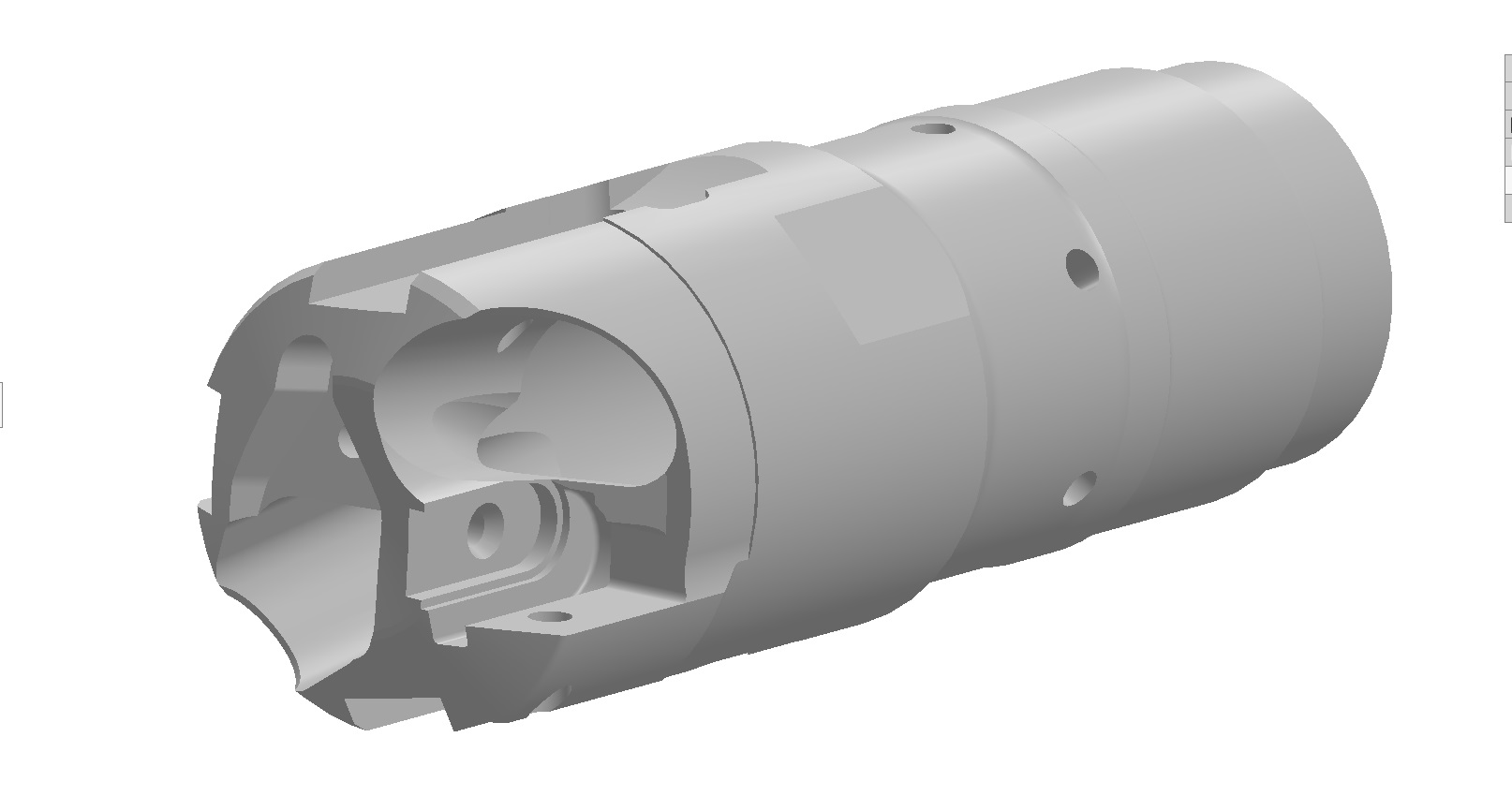}};
  \node[anchor=south, xshift=-\wframe, yshift = \hframe] at (center) {\includegraphics[trim=115 45 375 100, clip, height=\himage]{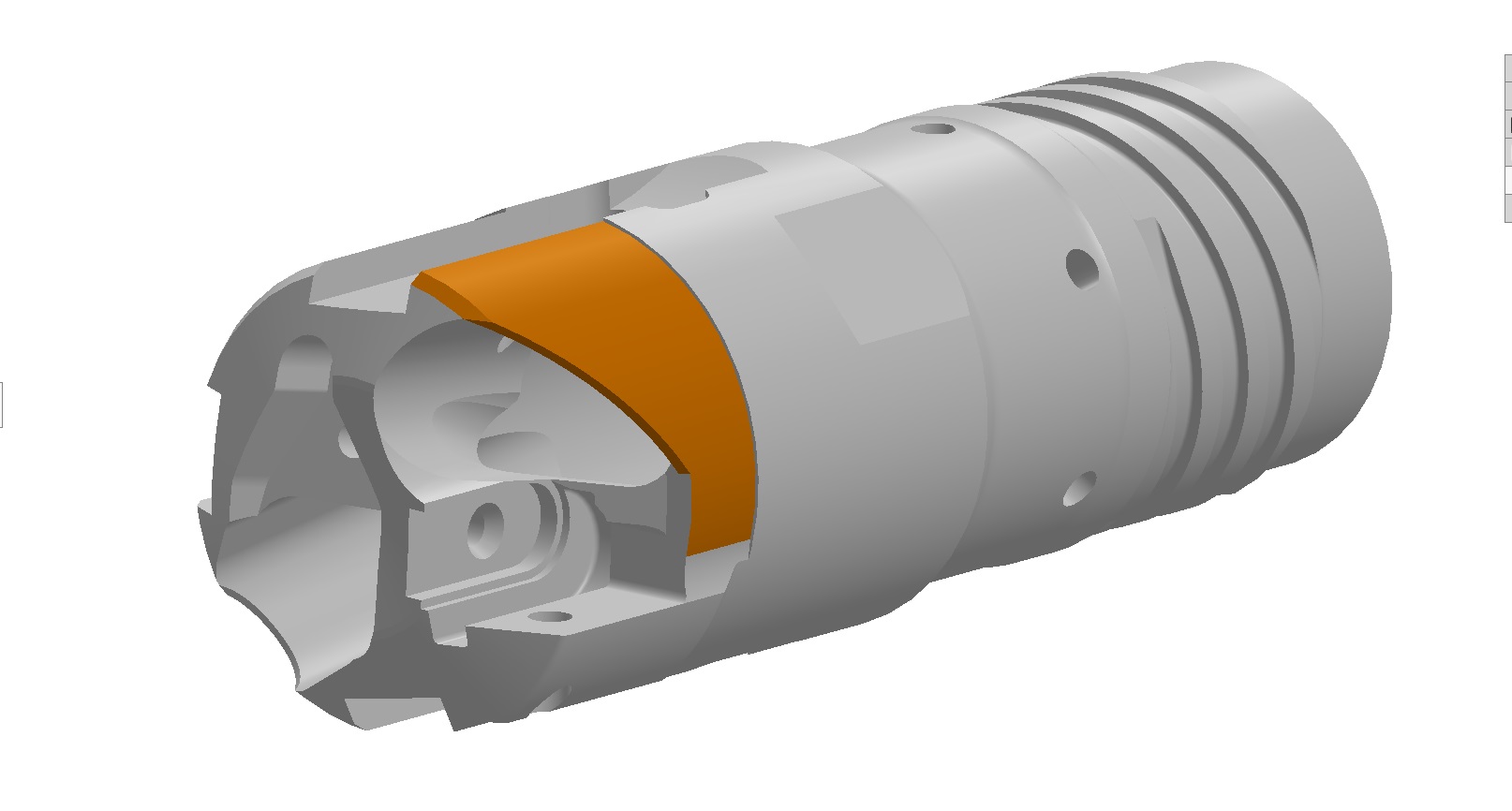}};
  \node[anchor=south west, xshift=-\wframe, yshift = \hframe] at (center_right) {\includegraphics[trim=115 45 375 100, clip, height=\himage]{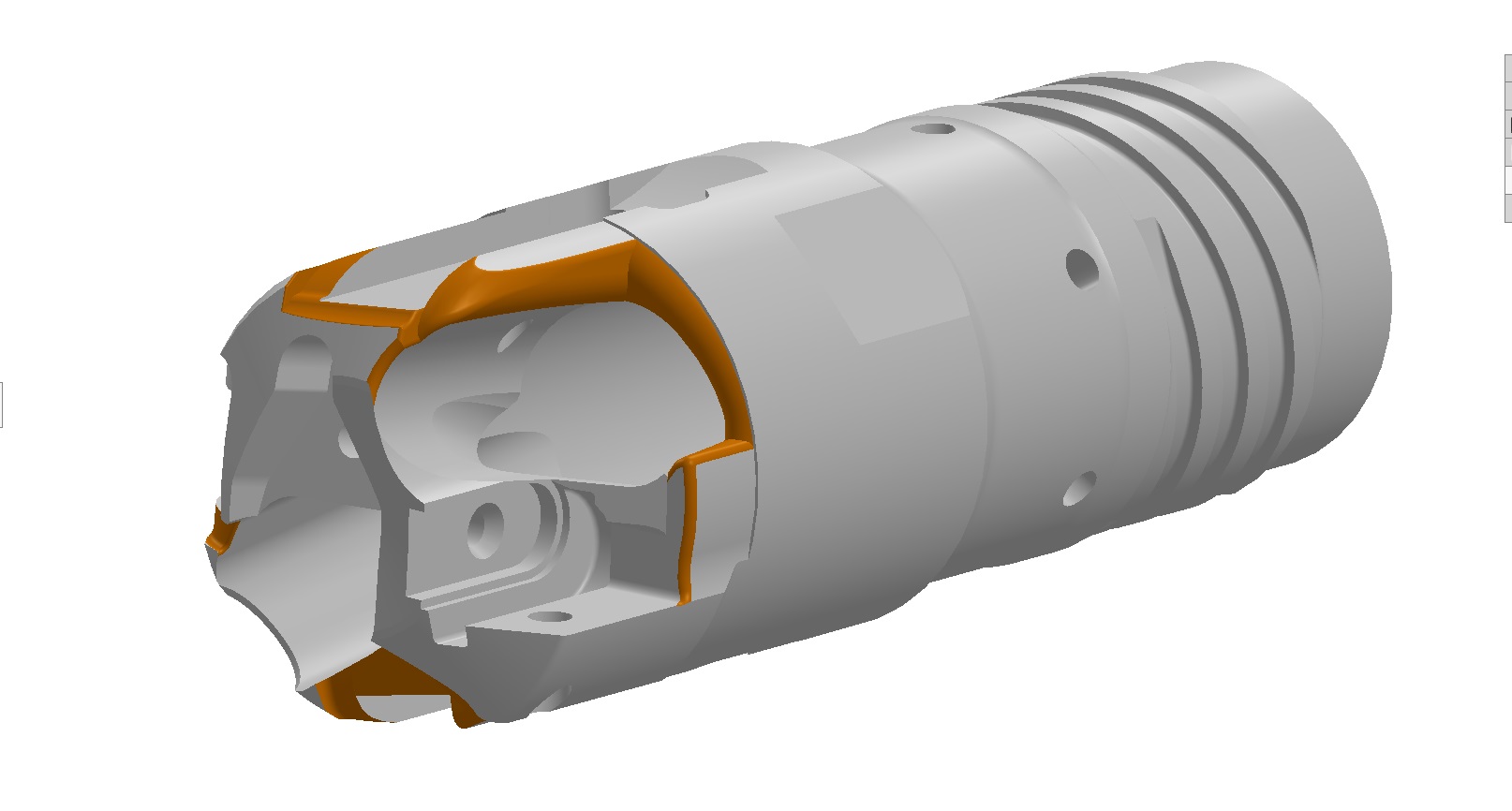}};

  % Desc
  \node[anchor=south east, xshift=-\wframe, yshift = \hframe] at (-1.65*\wimage + 1cm, 3.5cm) {reference};
  %\node[anchor=south west, xshift=-\wframe, yshift = \hframe] at (-1.5*\wimage, 3cm) {modification I};
  \node[anchor=south west, xshift=-\wframe, yshift = \hframe] at (-0.5*\wimage, 3.5cm) {modification II};
  %\node[anchor=south west, xshift=-\wframe, yshift = \hframe] at (0.5*\wimage, 3cm) {modification III};
  \node[anchor=south west, xshift=-\wframe, yshift = \hframe] at (1.0*\wimage, 3.5cm) {modification IV};

  % \draw[step=1cm,gray,very thin] (-4,0) grid (10,4);

\end{tikzpicture}
    \caption{Reference drill head and the modified designs with changes to the chip mouth highlighted according to \cite{Baumann2024}.}
    \label{fig:drills_overview}
\end{figure*}
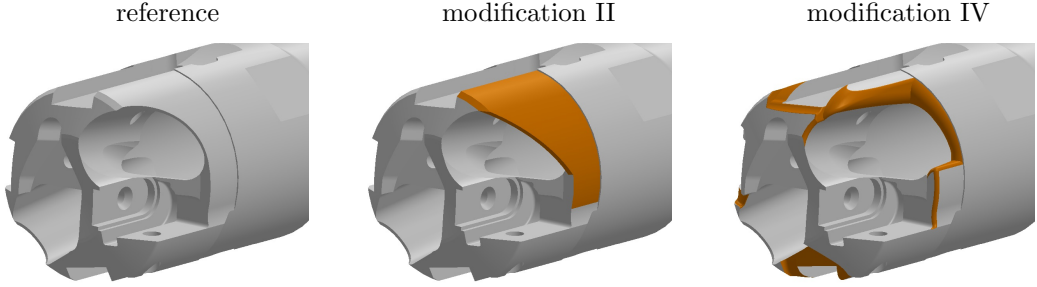

With the use of the SPH fundamentals described in the previous section, simulations are set up to investigate the chip-fluid interaction inside the ejector drill head and to compare the subsequent chip evacuation performance characteristics of the modified drill heads with respect to the reference drill head. Figure~\ref{fig:drills_overview} depicts the modified drill head designs investigated next to the reference drill head. The figure focuses on the chip mouth with the outer cutting edge. Modification~II aims to decrease the flow area near the cutting edge by reshaping the outer tool wall, increasing the coverage of the chip mouth to mitigate vortex formation at the chip mouth. Conversely, modification~IV  eliminates the outer wall to expand the flow area and improve the efficiency of chip evacuation based on faster flow rates~\cite{Baumann2024}.

The simulation starts with an empty bore hole and introduces a constant coolant inflow near the drill head to establish a quasi steady state at the cutting front. The inflow is modeled as laminar with a constant velocity of $5\,\textrm{m}\,\textrm{s}^{-1}$ and is positioned after the MWF outlet bore holes based on experimental Particle Image Velocimetry (PIV) data~\cite{Gerken2022}. The simulation includes a fluid rotation that matches the tool and the rotating outlet bores to replicate the actual conditions. Homogeneous properties of a ten percent concentrated metalworking fluid and water mixture is assumed  with effective properties calculated from the mixture ratio. Furthermore, isothermal conditions are applied, reflecting the experimental scenario in which the fluid temperature remains stable due to a $1{,}000\,\mathrm{liter}$ reservoir of MWF held at a temperature of $20\,^{\circ}\mathrm{C}$.

\begin{figure}[htb]

  \def\svgwidth{\columnwidth}
  
  \begingroup%
  \makeatletter%
  \providecommand\color[2][]{%
    \errmessage{(Inkscape) Color is used for the text in Inkscape, but the package 'color.sty' is not loaded}%
    \renewcommand\color[2][]{}%
  }%
  \providecommand\transparent[1]{%
    \errmessage{(Inkscape) Transparency is used (non-zero) for the text in Inkscape, but the package 'transparent.sty' is not loaded}%
    \renewcommand\transparent[1]{}%
  }%
  \providecommand\rotatebox[2]{#2}%
  
  %text size
  \footnotesize

  \newcommand*\fsize{\dimexpr\f@size pt\relax}%
  \newcommand*\lineheight[1]{\fontsize{\fsize}{#1\fsize}\selectfont}%
  \ifx\svgwidth\undefined%
    \setlength{\unitlength}{972bp}%
    \ifx\svgscale\undefined%
      \relax%
    \else%
      \setlength{\unitlength}{\unitlength * \real{\svgscale}}%
    \fi%
  \else%
    \setlength{\unitlength}{\svgwidth}%
  \fi%
  \global\let\svgwidth\undefined%
  \global\let\svgscale\undefined%
  \makeatother%
  \begin{picture}(1,0.80740743)%
    \lineheight{1}%
    \setlength\tabcolsep{0pt}%
    \put(0,0){\includegraphics[width=\unitlength,page=1]{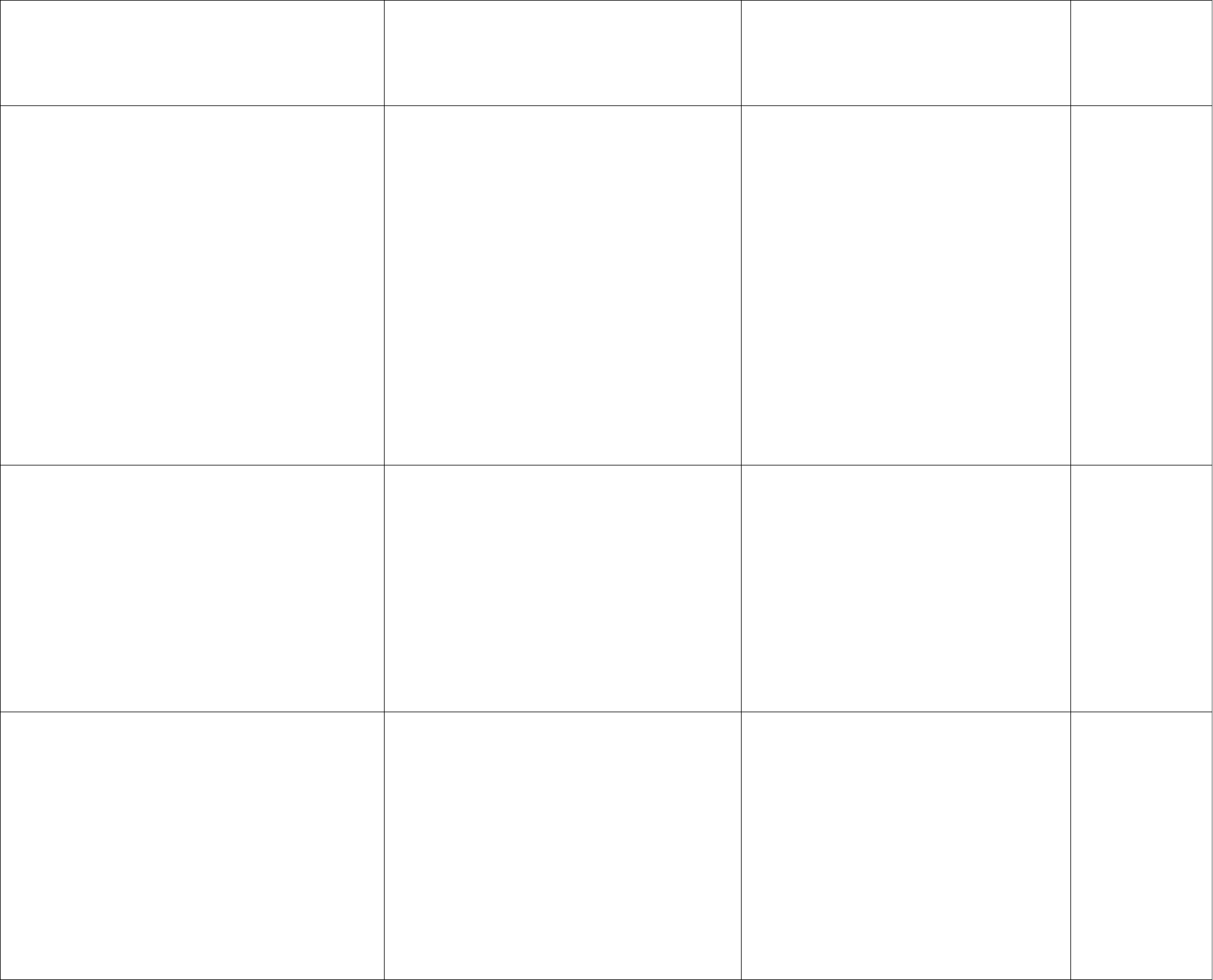}}%
    \put(0.005,0.75177969){\makebox(0,0)[lt]{\lineheight{1.25}\smash{\begin{tabular}[t]{l}center cutting edge\end{tabular}}}}%
    \put(0.35,0.77153278){\makebox(0,0)[lt]{\lineheight{1.25}\smash{\begin{tabular}[t]{l}inside of outer \end{tabular}}}}%
    \put(0.37,0.7320266){\makebox(0,0)[lt]{\lineheight{1.25}\smash{\begin{tabular}[t]{l}cutting edge \end{tabular}}}}%
    \put(0.62,0.77153278){\makebox(0,0)[lt]{\lineheight{1.25}\smash{\begin{tabular}[t]{l}outside of outer \end{tabular}}}}%
    \put(0.63,0.7320266){\makebox(0,0)[lt]{\lineheight{1.25}\smash{\begin{tabular}[t]{l}cutting edge\end{tabular}}}}%
    \put(0.07012346,0.48160479){\makebox(0,0)[lt]{\lineheight{1.25}\smash{\begin{tabular}[t]{l}\textit{n}\end{tabular}}}}%
    \put(0.08666667,0.47345664){\makebox(0,0)[lt]{\lineheight{1.25}\smash{\begin{tabular}[t]{l}\textit{s,z}\end{tabular}}}}%
    \put(0.13,0.48){\makebox(0,0)[lt]{\lineheight{1.25}\smash{\begin{tabular}[t]{l}= 0.72\end{tabular}}}}%
    \put(0.07012346,0.43592578){\makebox(0,0)[lt]{\lineheight{1.25}\smash{\begin{tabular}[t]{l}\textit{f}\end{tabular}}}}%
    \put(0.07925926,0.42775705){\makebox(0,0)[lt]{\lineheight{1.25}\smash{\begin{tabular}[t]{l}\textit{s,z}\end{tabular}}}}%
    \put(0.13,0.43592578){\makebox(0,0)[lt]{\lineheight{1.25}\smash{\begin{tabular}[t]{l}= 7.6 \end{tabular}}}}%
    \put(0.225,0.43592578){\makebox(0,0)[lt]{\lineheight{1.25}\smash{\begin{tabular}[t]{l}s\end{tabular}}}}%
    \put(0.238,0.44580232){\makebox(0,0)[lt]{\lineheight{1.25}\smash{\begin{tabular}[t]{l}-\end{tabular}}}}%
    \put(0.245,0.44580232){\makebox(0,0)[lt]{\lineheight{1.25}\smash{\begin{tabular}[t]{l}1\end{tabular}}}}%
    \put(0.64,0.48888874){\makebox(0,0)[lt]{\lineheight{1.25}\smash{\begin{tabular}[t]{l}\textit{n}\end{tabular}}}}%
    \put(0.66,0.48074059){\makebox(0,0)[lt]{\lineheight{1.25}\smash{\begin{tabular}[t]{l}\textit{s,aa}\end{tabular}}}}%
    \put(0.73,0.48888874){\makebox(0,0)[lt]{\lineheight{1.25}\smash{\begin{tabular}[t]{l}= 4.11\end{tabular}}}}%
    \put(0.64,0.44318915){\makebox(0,0)[lt]{\lineheight{1.25}\smash{\begin{tabular}[t]{l}\textit{f}\end{tabular}}}}%
    \put(0.65,0.435041){\makebox(0,0)[lt]{\lineheight{1.25}\smash{\begin{tabular}[t]{l}\textit{s,aa}\end{tabular}}}}%
    %\put(0.66384774,0.435041){\makebox(0,0)[lt]{\lineheight{1.25}\smash{\begin{tabular}[t]{l}\textit{,aa}\end{tabular}}}}%
    \put(0.721,0.44318915){\makebox(0,0)[lt]{\lineheight{1.25}\smash{\begin{tabular}[t]{l}= 43.67 \end{tabular}}}}%
    \put(0.848,0.44318915){\makebox(0,0)[lt]{\lineheight{1.25}\smash{\begin{tabular}[t]{l}s\end{tabular}}}}%
    \put(0.861,0.45306569){\makebox(0,0)[lt]{\lineheight{1.25}\smash{\begin{tabular}[t]{l}-\end{tabular}}}}%
    \put(0.868,0.45306569){\makebox(0,0)[lt]{\lineheight{1.25}\smash{\begin{tabular}[t]{l}1\end{tabular}}}}%
    \put(0.36763374,0.48301426){\makebox(0,0)[lt]{\lineheight{1.25}\smash{\begin{tabular}[t]{l}\textit{n}\end{tabular}}}}%
    \put(0.38417695,0.47486611){\makebox(0,0)[lt]{\lineheight{1.25}\smash{\begin{tabular}[t]{l}\textit{s,ai}\end{tabular}}}}%
    \put(0.44,0.48301426){\makebox(0,0)[lt]{\lineheight{1.25}\smash{\begin{tabular}[t]{l}= 3.11\end{tabular}}}}%
    \put(0.36763374,0.43733525){\makebox(0,0)[lt]{\lineheight{1.25}\smash{\begin{tabular}[t]{l}\textit{f}\end{tabular}}}}%
    \put(0.37676954,0.4291871){\makebox(0,0)[lt]{\lineheight{1.25}\smash{\begin{tabular}[t]{l}\textit{s,ai}\end{tabular}}}}%
    \put(0.44,0.43733525){\makebox(0,0)[lt]{\lineheight{1.25}\smash{\begin{tabular}[t]{l}= 32.97\end{tabular}}}}%
    \put(0.565,0.43733525){\makebox(0,0)[lt]{\lineheight{1.25}\smash{\begin{tabular}[t]{l}s\end{tabular}}}}%
    \put(0.578,0.44721179){\makebox(0,0)[lt]{\lineheight{1.25}\smash{\begin{tabular}[t]{l}-\end{tabular}}}}%
    \put(0.585,0.44721179){\makebox(0,0)[lt]{\lineheight{1.25}\smash{\begin{tabular}[t]{l}1\end{tabular}}}}%

    \put(0,0){\includegraphics[width=\unitlength,page=2]{Fig2.pdf}}%
    \put(0.20,0.239){\makebox(0,0)[lt]{\lineheight{1.25}\smash{\begin{tabular}[t]{l}4 mm\end{tabular}}}}%
    \put(0,0){\includegraphics[width=\unitlength,page=3]{Fig2.pdf}}%
    \put(0.51,0.239){\makebox(0,0)[lt]{\lineheight{1.25}\smash{\begin{tabular}[t]{l}4 mm\end{tabular}}}}%
    \put(0,0){\includegraphics[width=\unitlength,page=4]{Fig2.pdf}}%
    \put(0.78,0.239){\makebox(0,0)[lt]{\lineheight{1.25}\smash{\begin{tabular}[t]{l}4 mm\end{tabular}}}}%
    \put(0,0){\includegraphics[width=\unitlength,page=5]{Fig2.pdf}}%
    \put(0.79,0.662){\makebox(0,0)[lt]{\lineheight{1.25}\smash{\begin{tabular}[t]{l}5 mm\end{tabular}}}}%
    \put(0.51,0.662){\makebox(0,0)[lt]{\lineheight{1.25}\smash{\begin{tabular}[t]{l}5 mm\end{tabular}}}}%
    \put(0,0){\includegraphics[width=\unitlength,page=6]{Fig2.pdf}}%
    \put(0.22,0.662){\makebox(0,0)[lt]{\lineheight{1.25}\smash{\begin{tabular}[t]{l}5 mm\end{tabular}}}}%
    \put(0,0){\includegraphics[width=\unitlength,page=7]{Fig2.pdf}}%
    \put(0.93,0.43){\rotatebox{90.00000252}{\makebox(0,0)[lt]{\lineheight{1.25}\smash{\begin{tabular}[t]{l}experimental chip \end{tabular}}}}}%
    \put(0.97,0.47167681){\rotatebox{90.00000252}{\makebox(0,0)[lt]{\lineheight{1.25}\smash{\begin{tabular}[t]{l}morphologies \end{tabular}}}}}%
    \put(0.93,0.23){\rotatebox{90.00000252}{\makebox(0,0)[lt]{\lineheight{1.25}\smash{\begin{tabular}[t]{l}scanned 3D \end{tabular}}}}}%
    \put(0.97,0.23982495){\rotatebox{90.00000252}{\makebox(0,0)[lt]{\lineheight{1.25}\smash{\begin{tabular}[t]{l}model\end{tabular}}}}}%
    \put(0.93,0.015){\rotatebox{90.00000252}{\makebox(0,0)[lt]{\lineheight{1.25}\smash{\begin{tabular}[t]{l}smoothened \end{tabular}}}}}%
    \put(0.97,0.02){\rotatebox{90.00000252}{\makebox(0,0)[lt]{\lineheight{1.25}\smash{\begin{tabular}[t]{l}STL model\end{tabular}}}}}%
    \put(0,0){\includegraphics[width=\unitlength,page=8]{Fig2.pdf}}%
    \put(0.25,0.11772619){\makebox(0,0)[lt]{\lineheight{1.25}\smash{\begin{tabular}[t]{l}\textit{l}\end{tabular}}}}%
    \put(0.26,0.11019532){\makebox(0,0)[lt]{\lineheight{1.25}\smash{\begin{tabular}[t]{l}\textit{c,z}\end{tabular}}}}%
    %\put(0.28313786,0.11019532){\makebox(0,0)[lt]{\lineheight{1.25}\smash{\begin{tabular}[t]{l},\end{tabular}}}}%
    %\put(0.28869342,0.11019532){\makebox(0,0)[lt]{\lineheight{1.25}\smash{\begin{tabular}[t]{l}��\end{tabular}}}}%
    \put(0,0){\includegraphics[width=\unitlength,page=9]{Fig2.pdf}}%
    \put(0.535,0.12430026){\makebox(0,0)[lt]{\lineheight{1.25}\smash{\begin{tabular}[t]{l}\textit{l}\end{tabular}}}}%
    \put(0.545,0.11673853){\makebox(0,0)[lt]{\lineheight{1.25}\smash{\begin{tabular}[t]{l}\textit{c,ai}\end{tabular}}}}%
    %\put(0.56222222,0.11673853){\makebox(0,0)[lt]{\lineheight{1.25}\smash{\begin{tabular}[t]{l},\end{tabular}}}}%
    %\put(0.56777778,0.11673853){\makebox(0,0)[lt]{\lineheight{1.25}\smash{\begin{tabular}[t]{l}����\end{tabular}}}}%
    \put(0,0){\includegraphics[width=\unitlength,page=10]{Fig2.pdf}}%
    \put(0.795,0.12185171){\makebox(0,0)[lt]{\lineheight{1.25}\smash{\begin{tabular}[t]{l}\textit{l}\end{tabular}}}}%
    \put(0.805,0.11432084){\makebox(0,0)[lt]{\lineheight{1.25}\smash{\begin{tabular}[t]{l}\textit{c,aa}\end{tabular}}}}%
    %\put(0.8391358,0.11432084){\makebox(0,0)[lt]{\lineheight{1.25}\smash{\begin{tabular}[t]{l},\end{tabular}}}}%
    %\put(0.84469136,0.11432084){\makebox(0,0)[lt]{\lineheight{1.25}\smash{\begin{tabular}[t]{l}����\end{tabular}}}}%
    \put(0.04046914,0.03538257){\makebox(0,0)[lt]{\lineheight{1.25}\smash{\begin{tabular}[t]{l}center chip\end{tabular}}}}%
    \put(0.38432099,0.03538257){\makebox(0,0)[lt]{\lineheight{1.25}\smash{\begin{tabular}[t]{l}inner chip\end{tabular}}}}%
    \put(0.6568107,0.03392578){\makebox(0,0)[lt]{\lineheight{1.25}\smash{\begin{tabular}[t]{l}outer chip\end{tabular}}}}%
  \end{picture}%
\endgroup%

  \caption{Resulting chip shapes from the cutting edges and final STL models for the simulations \cite{Gerken2024}.}
  \label{fig:chip_shapes}
\end{figure}

Because a coupled simulation of chip formation as well as chip evacuation is computationally too expensive to model the highly sophisticated nature of the real process accurately, chips are introduced as fully formed rigid bodies into the simulation. The resulting chip shapes, their formation frequencies, and their positioning during the ejector drilling process are obtained by the extensive experimental tests explained in~\cite{Gerken2024}, which are vital to set up an accurate simulation model. As shown in Figure~\ref{fig:chip_shapes}, several types of the chips formed from the two cutting edges are identified as well as their positioning during the drilling process. The center chip is formed by the center cutting edge. Inner and outer chips are formed concurrently by the outer cutting edge due to its shape. However, chip forming frequency $\mathrm{f}_{s}$ and number of chips generated per tool rotation $\mathrm{n}_{s}$ are different for these three chip shapes. Although the chip shape profile, $\mathrm{f}_{s}$ and $\mathrm{n}_{s}$ vary depending on the process parameters such as cutting speed $\mathrm{v}_{c}$, feed rate $\mathrm{f}$, and volume flow rate $\mathrm{\dot{V}_{\Sigma}}$, a common reference chip form for each cutting edge is selected as the center chip, the inner chip, and the outer chip in the simulations. Additionally, smoothened versions with reduced mesh triangle resolution of the scanned and digitized reference chip forms are used to improve the efficiency of the simulation. The impact on the chip-fluid interactions by smoothening the chip shapes is assumed to be negligible.

\begin{figure}[htb]
 
  \def\svgwidth{\columnwidth}

  \begingroup%
  \makeatletter%
  \providecommand\color[2][]{%
    \errmessage{(Inkscape) Color is used for the text in Inkscape, but the package 'color.sty' is not loaded}%
    \renewcommand\color[2][]{}%
  }%
  \providecommand\transparent[1]{%
    \errmessage{(Inkscape) Transparency is used (non-zero) for the text in Inkscape, but the package 'transparent.sty' is not loaded}%
    \renewcommand\transparent[1]{}%
  }%
  \providecommand\rotatebox[2]{#2}%

  % Set font size
  \footnotesize
  
  \newcommand*\fsize{\dimexpr\f@size pt\relax}%
  \newcommand*\lineheight[1]{\fontsize{\fsize}{#1\fsize}\selectfont}%
  \ifx\svgwidth\undefined%
    \setlength{\unitlength}{572.26654293bp}%
    \ifx\svgscale\undefined%
      \relax%
    \else%
      \setlength{\unitlength}{\unitlength * \real{\svgscale}}%
    \fi%
  \else%
    \setlength{\unitlength}{\svgwidth}%
  \fi%
  \global\let\svgwidth\undefined%
  \global\let\svgscale\undefined%
  \makeatother%
  \begin{picture}(1,0.56858023)%
    \lineheight{1}%
    \setlength\tabcolsep{0pt}%
    \put(0,0){\includegraphics[width=\unitlength,page=1]{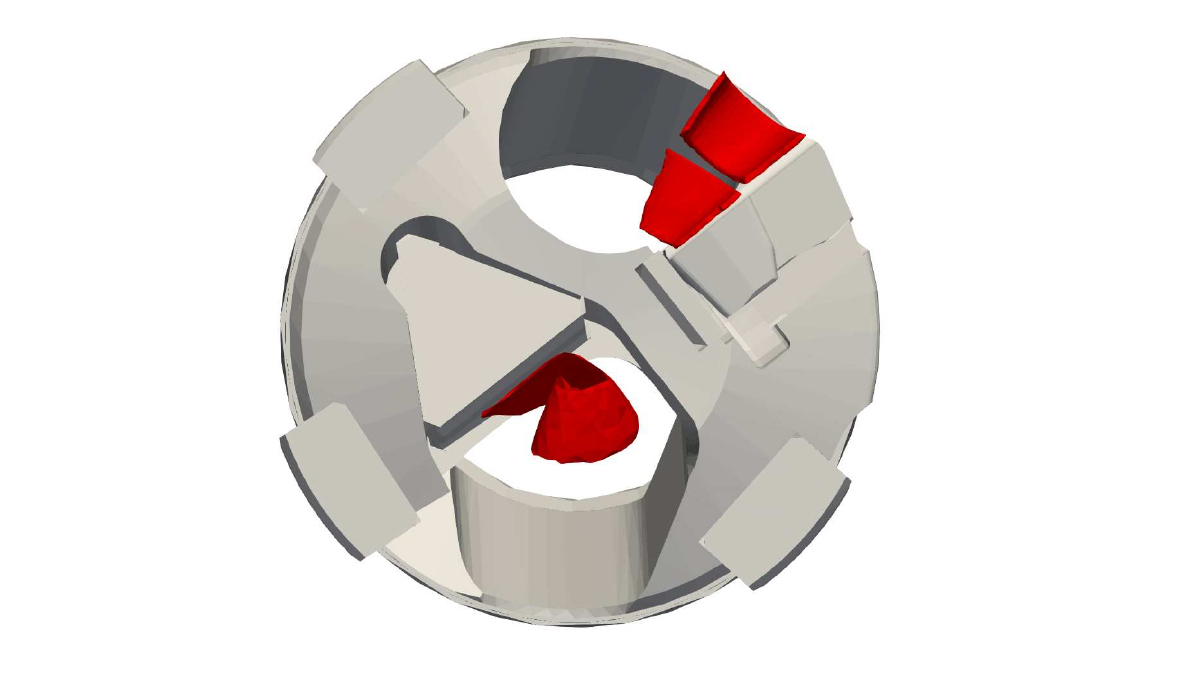}}%
    \put(0.67200869,0.324414){\makebox(0,0)[lt]{\lineheight{1.25}\smash{\begin{tabular}[t]{l}outer cutting edge\end{tabular}}}}%
    \put(0.02,0.51){\makebox(0,0)[lt]{\lineheight{1.25}\smash{\begin{tabular}[t]{l}inner cutting edge\end{tabular}}}}%
    \put(0.78,0.40){\makebox(0,0)[lt]{\lineheight{1.25}\smash{\begin{tabular}[t]{l}inner chip\end{tabular}}}}%
    \put(0.09,0.075){\makebox(0,0)[lt]{\lineheight{1.25}\smash{\begin{tabular}[t]{l}center chip\end{tabular}}}}%
    \put(0.785,0.475){\makebox(0,0)[lt]{\lineheight{1.25}\smash{\begin{tabular}[t]{l}outer chip\end{tabular}}}}%
    \put(0.75035093,0.18216212){\makebox(0,0)[lt]{\lineheight{1.25}\smash{\begin{tabular}[t]{l}chip mouths\end{tabular}}}}%
    \put(0,0){\includegraphics[width=\unitlength,page=2]{Fig3.pdf}}%
    \put(0.08789198,0.29971358){\makebox(0,0)[lt]{\lineheight{1.25}\smash{\begin{tabular}[t]{l}Y\end{tabular}}}}%
    \put(0.1577096,0.23515677){\makebox(0,0)[lt]{\lineheight{1.25}\smash{\begin{tabular}[t]{l}X\end{tabular}}}}%
    \put(0.08145497,0.22924946){\makebox(0,0)[lt]{\lineheight{1.25}\smash{\begin{tabular}[t]{l}Z\end{tabular}}}}%
  \end{picture}%
\endgroup%

  \caption{Chip positioning at $\textrm{t = 0.01\,s}$ for drill head top view.}
  \label{fig:chip positioning}
\end{figure}

As depicted in Figure~\ref{fig:chip positioning}, initially all three types of chips are positioned outside the bottom of the bore hole next to the cutting edges of the drill head at $\textrm{t} = 0.01\,\textrm{s}$ and are inserted at a prescribed longitudinal velocity $\textrm{v}_{chip,z}$. Similar to the real experimental scenario, they are then released to interact and move along with the fluid once they are completely inside the chip mouth. The chip velocities and positions are set relative to the rotating drill head frame until they are released during the insertion. Individual chip insert velocities for outer chip, inner chip, and center chip are derived using the chip forming frequency and chip's characteristic length $\textrm{l}_{c,z}$ according to Figure~\ref{fig:chip_shapes}. Chip lengths are measured and defined along the longitudinal axis of the drill head using CAD software. New chips are introduced progressively during the simulation with the respective characteristics once the previous chips are released in the chip mouth. To maintain numerical stability and to avoid excessive contact forces during insertion, the spacing between cutting edges and chips is carefully adjusted. Similarly, the time intervals of chip insertions between two similar chips are modified to ensure that each succeeding chip only begins interacting with the fluid once the preceding chip is fully released in the chip mouth.

 Since the cutting edges are the same for the reference drill head and modified drill heads, chip forms are assumed to be the same when operating all drill head variations. Table~\ref{tab:sph_parameters} lists the boundary conditions and input parameters applied for the coupled SPH-DEM  simulations. 

\begin{table}[ht!]
\caption{Simulation input parameters and boundary conditions.}
\label{tab:sph_parameters}
\centering
\begin{tabular}{@{}llll@{}}
\toprule
\textbf{property} & \textbf{symbol} & \textbf{value} \\
\midrule
drill diameter & $D$ & $30\,\mathrm{mm}$ \\
fluid density & $\rho$ & $999.3\,\mathrm{kg\,m^{-3}}$ \\
dynamic viscosity & $\mu$ & $0.01\,\mathrm{kg\,m^{-1}s^{-1}}$ \\
fluid inflow velocity & $v$ & $5\,\mathrm{m\,s^{-1}}$ \\
rotation of the drill & $\omega$ & $66.7\,\mathrm{rad\,s^{-1}}$ \\
artificial viscosity factors & $\alpha$& $0.05$ \\
    &$\beta$  &  $0.1$ \\
artificial stress factor & - & $0.2$ \\
diffuse density factor & $\delta$ & $0.1$ \\
artificial speed of sound & $c_o$ & $50\,\mathrm{m\,s^{-1}}$ \\
CFL number & $\alpha_c$ & $0.1$ \\
chip density & $\rho_{chip}$ & $7800\,\mathrm{kg\,m^{-3}}$ \\
initial particle distance & $\Delta$ & $3.5e-4\,\mathrm{m}$ \\
smoothing length & $h$ & $5.25e-4\,\mathrm{m}$ \\

\textbf{chip insert velocities}&  &  \\
outer chip & $v_{chip,z_1}$ & $0.14114\,\mathrm{m\,s^{-1}}$ \\
inner chip & $v_{chip,z_2}$ & $0.08843\,\mathrm{m\,s^{-1}}$ \\
center chip & $v_{chip,z_3}$ & $0.03991\,\mathrm{m\,s^{-1}}$ \\
\bottomrule
\end{tabular}
\end{table}

\maketitle
\section{Simulation Results and Discussion}\label{sec3}

\begin{figure*}[htb]
    \includegraphics[width=\textwidth,keepaspectratio]{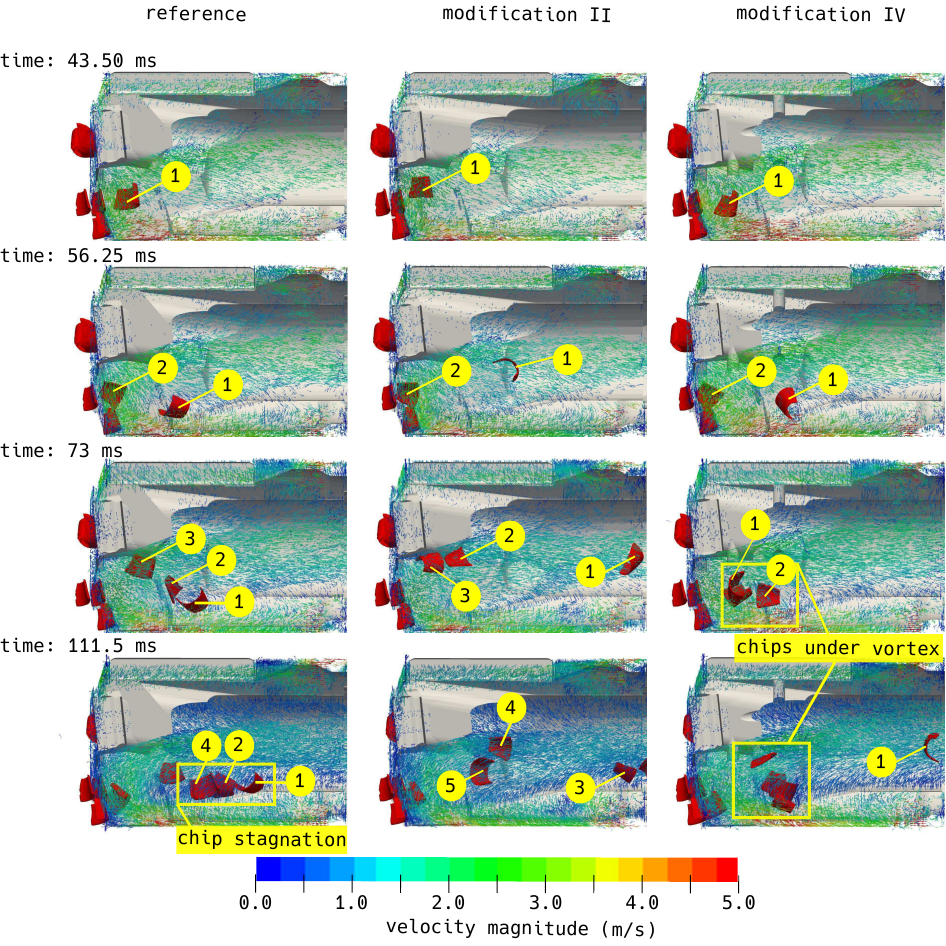}
    \caption{Positions of the inner and outer chips (numbered according to generation) during the transient phase of chip evacuation.}
    \label{fig:transient chip evacuation}
\end{figure*}

\begin{figure*}[htb]
    \centering
    \def\svgwidth{\textwidth}

    \begingroup%
  \makeatletter%
  \providecommand\color[2][]{%
    \errmessage{(Inkscape) Color is used for the text in Inkscape, but the package 'color.sty' is not loaded}%
    \renewcommand\color[2][]{}%
  }%
  \providecommand\transparent[1]{%
    \errmessage{(Inkscape) Transparency is used (non-zero) for the text in Inkscape, but the package 'transparent.sty' is not loaded}%
    \renewcommand\transparent[1]{}%
  }%
  \providecommand\rotatebox[2]{#2}%
  \newcommand*\fsize{\dimexpr\f@size pt\relax}%
  \newcommand*\lineheight[1]{\fontsize{\fsize}{#1\fsize}\selectfont}%
  \ifx\svgwidth\undefined%
    \setlength{\unitlength}{453.54330709bp}%
    \ifx\svgscale\undefined%
      \relax%
    \else%
      \setlength{\unitlength}{\unitlength * \real{\svgscale}}%
    \fi%
  \else%
    \setlength{\unitlength}{\svgwidth}%
  \fi%
  \global\let\svgwidth\undefined%
  \global\let\svgscale\undefined%
  \makeatother%
  \begin{picture}(1,1)%
    \lineheight{1}%
    \setlength\tabcolsep{0pt}%
    \put(-0.00105087,0.92279276){\rotatebox{-0.82603631}{\makebox(0,0)[lt]{\lineheight{1.25}\smash{\begin{tabular}[t]{l}reference\end{tabular}}}}}%
    \put(0,0){\includegraphics[width=\unitlength,page=1]{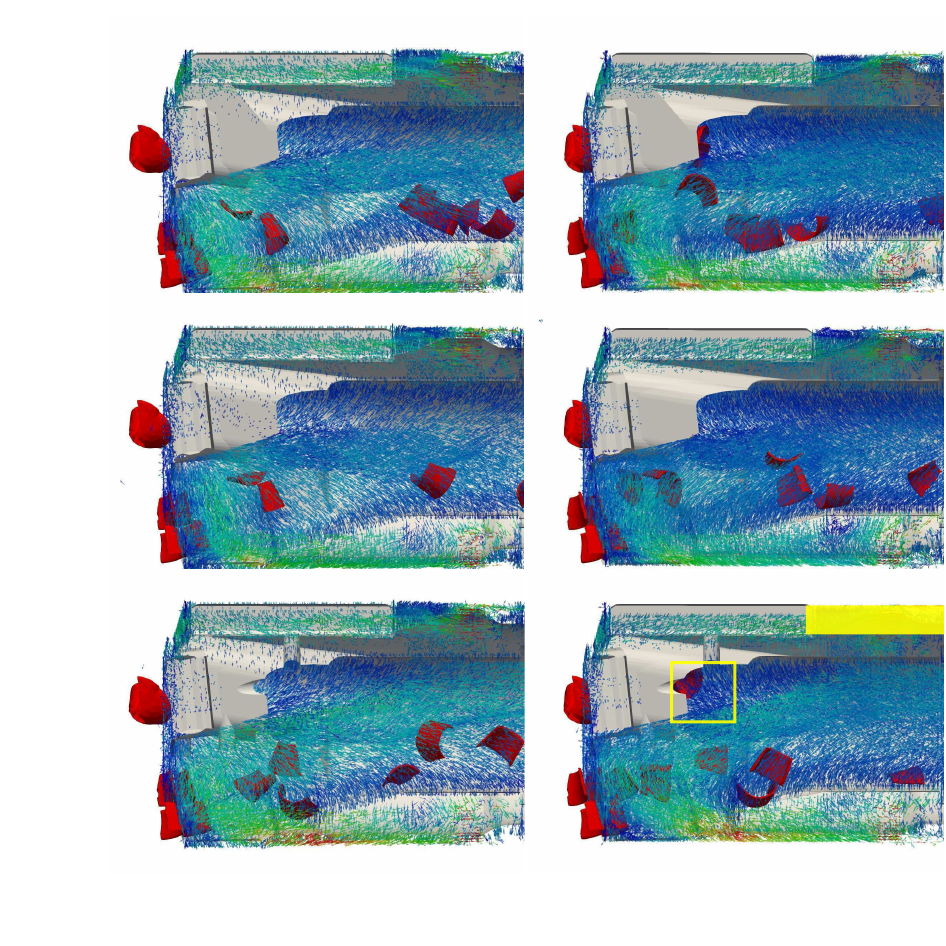}}%
    \put(0.86094689,0.34015594){\rotatebox{-0.82603631}{\makebox(0,0)[lt]{\lineheight{1.25}\smash{\begin{tabular}[t]{l}center chip\end{tabular}}}}}%
    \put(0,0){\includegraphics[width=\unitlength,page=2]{Fig6.pdf}}%
    \put(0.21746132,0.03866179){\makebox(0,0)[lt]{\lineheight{1.25}\smash{\begin{tabular}[t]{l}0.0\end{tabular}}}}%
    \put(0.37302341,0.01368402){\rotatebox{-0.82603631}{\makebox(0,0)[lt]{\lineheight{1.25}\smash{\begin{tabular}[t]{l}velocity magnitude (m/s)\end{tabular}}}}}%
    \put(0.31869631,0.03866179){\makebox(0,0)[lt]{\lineheight{1.25}\smash{\begin{tabular}[t]{l}1.0\end{tabular}}}}%
    \put(0.42151362,0.03866179){\makebox(0,0)[lt]{\lineheight{1.25}\smash{\begin{tabular}[t]{l}2.0\end{tabular}}}}%
    \put(0.52441628,0.03866179){\makebox(0,0)[lt]{\lineheight{1.25}\smash{\begin{tabular}[t]{l}3.0\end{tabular}}}}%
    \put(0.62654431,0.03866179){\makebox(0,0)[lt]{\lineheight{1.25}\smash{\begin{tabular}[t]{l}4.0\end{tabular}}}}%
    \put(0.72862752,0.03866179){\makebox(0,0)[lt]{\lineheight{1.25}\smash{\begin{tabular}[t]{l}5.0\end{tabular}}}}%
    \put(0.1862307,0.9661734){\makebox(0,0)[lt]{\lineheight{1.25}\smash{\begin{tabular}[t]{l}time: 155 ms\end{tabular}}}}%
    \put(0.62373072,0.9661734){\makebox(0,0)[lt]{\lineheight{1.25}\smash{\begin{tabular}[t]{l}time: 202 ms\end{tabular}}}}%
    \put(0.00081897,0.64153662){\rotatebox{-0.82603631}{\makebox(0,0)[lt]{\lineheight{1.25}\smash{\begin{tabular}[t]{l}modification II\end{tabular}}}}}%
    \put(0.00081897,0.34778658){\rotatebox{-0.82603631}{\makebox(0,0)[lt]{\lineheight{1.25}\smash{\begin{tabular}[t]{l}modification IV\end{tabular}}}}}%
    \put(0,0){\includegraphics[width=\unitlength,page=3]{Fig6.pdf}}%
    \put(0.86094915,0.92199888){\rotatebox{-0.82603631}{\makebox(0,0)[lt]{\lineheight{1.25}\smash{\begin{tabular}[t]{l}center chip\end{tabular}}}}}%
    \put(0,0){\includegraphics[width=\unitlength,page=4]{Fig6.pdf}}%
    \put(0.8171935,0.67236025){\rotatebox{-0.82603631}{\makebox(0,0)[lt]{\lineheight{1.25}\smash{\begin{tabular}[t]{l}chip stagnation\end{tabular}}}}}%
    \put(0,0){\includegraphics[width=\unitlength,page=5]{Fig6.pdf}}%
    \put(0.31176969,0.67226217){\rotatebox{-0.82603631}{\makebox(0,0)[lt]{\lineheight{1.25}\smash{\begin{tabular}[t]{l}initial chip cluster\end{tabular}}}}}%
    \put(0,0){\includegraphics[width=\unitlength,page=6]{Fig6.pdf}}%
  \end{picture}%
\endgroup%

    \caption{Positions of the inner and outer chips during quasi-steady state of chip evacuation.}
    \label{fig:quasi_steady chip evacuation}
\end{figure*}

The positions of the rigid chips and the velocity profiles of the MWF flow inside the drill heads at different time steps during the transient phase and quasi-steady state for the reference drill head and the modified drill heads resulting from coupled SPH-DEM simulations are depicted in Figures~\ref{fig:transient chip evacuation} and~\ref{fig:quasi_steady chip evacuation}, respectively. For each timestep, the reference drill head is located on the left, the modification design II is located on the center, and the modification design IV is at the right-hand side in the Figure~\ref{fig:transient chip evacuation}. The figure highlights the flow characteristics of the chip-fluid mixture from the outer cutting edge to the end of the drill head through the chip mouth and the internal geometry of the drill head. The fluid influx is positioned on the right side with velocity vectors visualizing the flow directions. The velocity magnitude of the fluid flow can be identified by the color legend. Inserted and released chips are shown in red to identify their positions and behavior at different time-steps throughout the simulation. Furthermore, chips are numbered according to the order of the generation.

The first outer chip and the first inner chip generated from the outer cutting edge are released at $\textrm{t} = 34.25\,\textrm{ms}$ and $\textrm{t} = 43.5\,\textrm{ms}$, respectively. By the time the first inner chip is released, the previously generated outer chip is already carried away from the cutting edge and into the chip mouth by the flow current. This behavior is consistently observed across all drill head models. In both, the reference design and modification II, the inner chip then stagnates near the cutting edge and remains in the cutting zone until approximately $\textrm{t} = 56.25\,\textrm{ms}$ without being fully evacuated. The rotation of the chip is mainly due to the velocity differences between the corners of the inner chip caused by the MWF flow current entering the drill tube from the edge of the chip mouth along the drill circumference. This rotational behavior is consistently seen in the subsequent inner chips for both, the reference design and modification II. In contrast, the duration of the inner chip stagnation near the cutting edge is shorter for modification IV. Although the chip rotates slightly, it is rapidly drawn towards the edge of the chip mouth because of the initial vortex formed near the internal edge of the chip mouth. This vortex is formed due to the strong turbulent flow near the edge of the chip mouth at the beginning, and it gradually diminishes over time as the drill head is completely filled with the MWF. 

As depicted at $\textrm{t} = 73\,\textrm{ms}$, the first outer chip that is inserted, reaches the end of the drill head for modification II. Conversely, it takes $\textrm{t} = 111.5\,\textrm{ms}$ for the first chip to reach the end of the drill head for modification IV, due to the initially generated vortex inside the chip mouth. Likewise for modification II, this is also the first outer chip that is inserted in the simulation. Further, it is observed that the third overall chip according to the order of chip generation becomes the first chip to reach the end of the drill head for the reference design. This takes place at $\textrm{t} = 100\,\textrm{ms}$. This is the second outer chip that is inserted in the simulation and this chip passes both the first outer and inner chips inside the drill head. This is due to the stagnation of the chips caused by the effects of the boundary layer near the internal wall of the reference drill head. This stagnation can still be seen at $\textrm{t} = 111.5\,\textrm{ms}$ as illustrated in Figure~\ref{fig:transient chip evacuation}. By this time, the third overall chip of modification II is already reaching the end of the drill head, demonstrating its stable and rapid chip evacuation behavior in the respective chip mouth from the outset.

Approximately at $\textrm{t} = 155\,\textrm{ms}$ the chip-MWF flow reaches a quasi-steady state in the drill heads as shown in Figure~\ref{fig:quasi_steady chip evacuation}. It is seen that the initially stagnated group of chips (which consists of four chips) gradually reaches the end of the reference drill head. By comparison, both modifications do not produce this type of stagnation due to boundary layer effects resulting from their improved internal flow. At $\textrm{t} = 202\,\textrm{ms}$, a new stagnation near the inner wall of the tube can be observed for the reference design. However, during the quasi-steady flow, evacuation of the released inner chips at the outer cutting edge is supported by and pushed towards the inward bound flow by the subsequently generated chip. This is due to weak and stagnating velocities of the MWF at the cutting front for all designs. Furthermore, the center spiral chip has already reached the internal tube of the drill head via chip mouth at $\textrm{t} = 202\,\textrm{ms}$ for both, modification IV and reference drill head. In contrast, this is not observed in modification II, exhibiting an inefficient weak flow near the cutting edge and in the chip mouth.

Table~\ref{tab:velocity} details the approximate velocity magnitudes as a range in the key regions of the drill head for all designs after the quasi-steady state is reached. As shown in~\cite{Baumann2024}, lubricant-based simulations show that modification IV with the extended chip mouth opening produces a higher lubricant flow rate both around the drill head and the outer cutting edge. The resulting effects of this are further exhibited and validated within the drill head for modification IV, as depicted by the velocity profiles in Figure~\ref{fig:quasi_steady chip evacuation}. Here, it is seen that modification IV maintains a strong flow with a higher velocity magnitude across the internal tube of the drill head between $0.5\cdots\,2\,\textrm{m}\,\textrm{s}^{-1}$ compared to the reference design and modification II, which are in the range of approximately $0.25\cdots\,1\,\textrm{m}\,\textrm{s}^{-1}$. Additionally, the velocity inside the chip mouth ranges between $1\cdots\,3\,\textrm{m}\,\textrm{s}^{-1}$ compared to $1\cdots\,2.5\,\textrm{m}\,\textrm{s}^{-1}$ for modification II. This is caused by the expanded flow area by removing the outer wall of the chip mouth in modification IV. However, it is observed that the initial inflow velocity of $5\,\textrm{m}\,\textrm{s}^{-1}$ of the MWF gets slowed down to a velocity magnitude of approximately $1\,\textrm{m}\,\textrm{s}^{-1}$ at the cutting edges for all the designs.

\begin{table}[ht!]
\caption{Velocity magnitudes at main regions of the drill head designs.}
\label{tab:velocity}
\centering
\renewcommand{\arraystretch}{1.25} % Optional: more vertical space
\begin{tabularx}{\columnwidth}{>{\hsize=1.05\hsize}X >{\centering\arraybackslash}X >{\centering\arraybackslash}X >{\centering\arraybackslash}X}
\toprule
\textbf{area} & \multicolumn{3}{c}{\textbf{velocity magnitude range}} \\
%\cmidrule(lr){2-4}
\cdashline{2-4}
& reference design & modification II & modification IV \\
\midrule
\shortstack{\rule{0pt}{8pt}chip\\ mouth} & \shortstack[l]{$1\cdots$\\$3\,\textrm{m}\,\textrm{s}^{-1}$} & \shortstack[l]{$1\cdots$\\$2.5\,\textrm{m}\,\textrm{s}^{-1}$} & \shortstack[l]{$1\cdots$\\$3\,\textrm{m}\,\textrm{s}^{-1}$} \\

\cdashline{1-4}
\shortstack{\rule{0pt}{12pt}inside of \\drill head } & \shortstack[l]{$0.25\cdots$\\$1\,\textrm{m}\,\textrm{s}^{-1}$} & \shortstack[l]{$0.25\cdots$\\$1\,\textrm{m}\,\textrm{s}^{-1}$} & \shortstack[l]{$0.5\cdots$\\$2\,\textrm{m}\,\textrm{s}^{-1}$} \\
\bottomrule

\end{tabularx}
\end{table}

Using the results from the overall simulation for modification II, it is observed that the flow current has become more streamlined and moved inward inside the modified drill head, improving the efficiency and supporting the chip evacuation process compared to the reference design. Although the velocity magnitudes are not higher than in the reference model, the boundary layer effects on chips in the internal tube of the drill head are not found for modification II. The stagnation of the chips near the internal tube wall due to the boundary layer effect is prominent in the reference design. As expected from modification II described in~\cite{Baumann2024}, the reduction in turbulence and vorticity in the fluid has accelerated the chip evacuation from the cutting zone to the drill tube.

However, although it is observed that modification II exhibits a stable chip evacuation behavior, it can be expected that modification IV outputs better and more efficient performance in longer durations since it maintains a much higher flow rate inside the drill head throughout the simulation. Modification IV has moved the generated vortex away from the cutting front to the back of the chip mouth due to it's extended chip mouth. This in turn supports the removal of chip blockage at the cutting front. Although modification II has reduced the vortex, it still remains at the cutting front.

Since the investigated designs are under the same constant inflow velocity, chip evacuation gains observed in the modifications result solely from the described chip mouth modifications. During the actual drilling process, the ejector effect is already established before the material removal process begins. This ejector effect further supports the chip evacuation behavior at the cutting front due to the supportive negative pressure generated on the lubricant inside the inner tube.  

Due to the large computational time durations of the simulations, the sample size for the center chips is not sufficient to understand the complete behavior of spiral chip evacuation. Since center chips have a very small chip formation frequency, only one chip evacuation from the cutting zone is observed until $\textrm{t} = 202\,\textrm{ms}$. Furthermore, as illustrated in Figure~\ref{fig:chip_shapes}, simulations are focused only on the three main reference chip shapes for each cutting edge without considering changes in the chip shape depending on the process parameters. However, during the real drilling process the chip shapes are not always identical and belt chips discussed in \cite{Gerken2024} are prone to cause more chip clogging in the chip mouths of the ejector drill head. In these simulations, thermal effects that are generated during the drilling process are omitted as well. Therefore, these results produce a conservative yet reliable picture about the chip evacuation behavior with regards to flow changes resulting from various design modifications to the ejector drill head.

\maketitle
\section{Experimental Investigations}\label{sec4}

As part of the comprehensive simulative analyses of various drill head designs using the SPH simulation, two designs were identified that increase the efficiency of the ejector drilling process either due to reduced vortex formation through a more closed chip mouth (modification II) or by increasing the local flow velocity by enlarging the chip mouth (modification IV). In order to validate the results of the simulations and to realize an efficient ejector drilling process, drill heads were produced using an additive manufacturing process. The influence on the MWF flow and chip removal was analyzed in experimental tests compared to the reference drill head without modifications. The most important criterion in the investigations is the determination of the minimum volume flow of the MWF supply at which a stable drilling process without chip blockage can be realized with the respective drill heads.

\subsection{Manufacturing of Modified Drill Heads}\label{subsec1}

\begin{figure}[htb]
     \def\svgwidth{\columnwidth}

    \begingroup%
  \makeatletter%
  \providecommand\color[2][]{%
    \errmessage{(Inkscape) Color is used for the text in Inkscape, but the package 'color.sty' is not loaded}%
    \renewcommand\color[2][]{}%
  }%
  \providecommand\transparent[1]{%
    \errmessage{(Inkscape) Transparency is used (non-zero) for the text in Inkscape, but the package 'transparent.sty' is not loaded}%
    \renewcommand\transparent[1]{}%
  }%
  \providecommand\rotatebox[2]{#2}%
 % Set font size
  \small
  
  \newcommand*\fsize{\dimexpr\f@size pt\relax}%
  \newcommand*\lineheight[1]{\fontsize{\fsize}{#1\fsize}\selectfont}%
  \ifx\svgwidth\undefined%
    \setlength{\unitlength}{746.25bp}%
    \ifx\svgscale\undefined%
      \relax%
    \else%
      \setlength{\unitlength}{\unitlength * \real{\svgscale}}%
    \fi%
  \else%
    \setlength{\unitlength}{\svgwidth}%
  \fi%
  \global\let\svgwidth\undefined%
  \global\let\svgscale\undefined%
  \makeatother%
  \begin{picture}(1,1.24321608)%
    \lineheight{1}%
    \setlength\tabcolsep{0pt}%
    \put(0,0){\includegraphics[width=\unitlength,page=1]{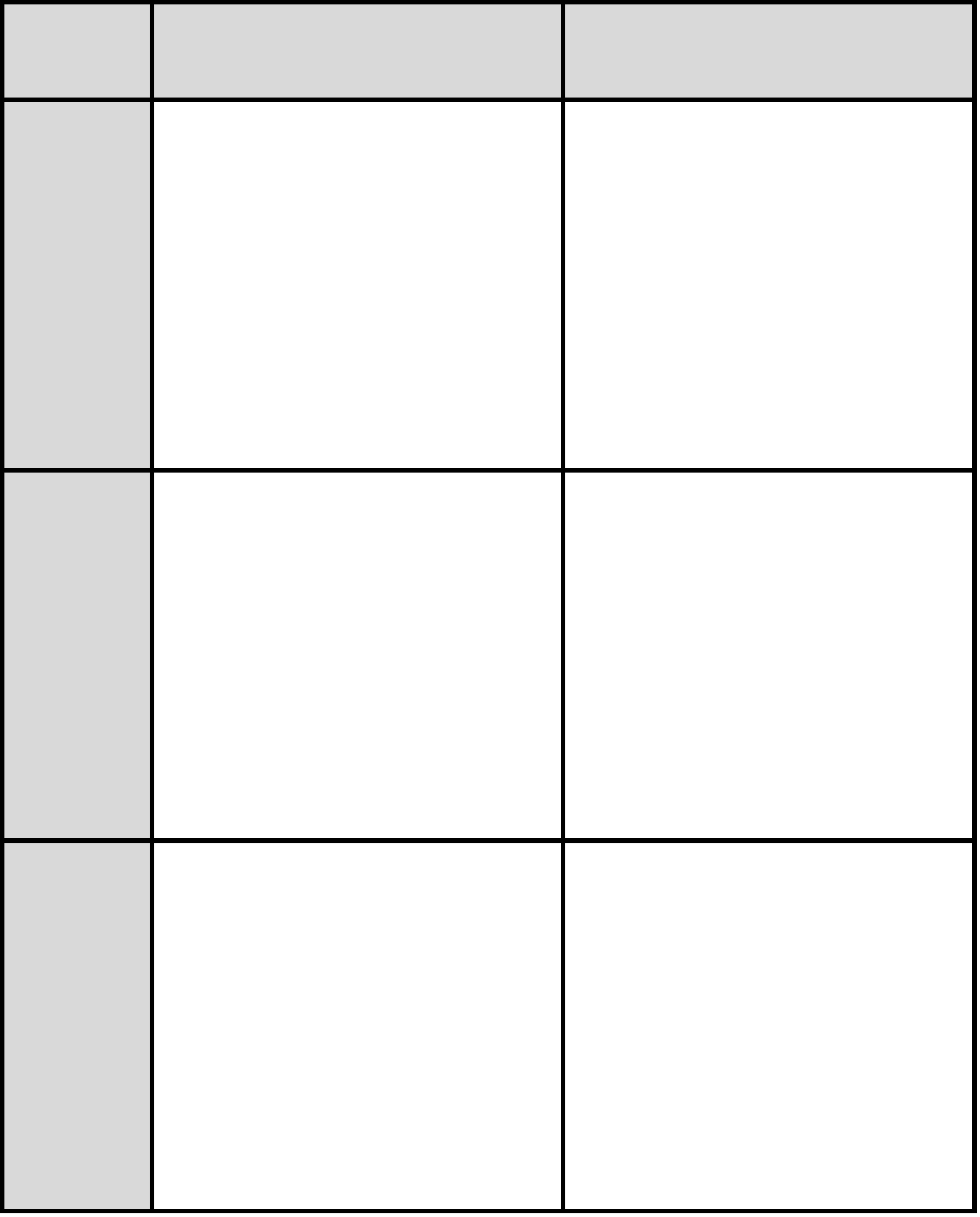}}%
    \put(0.16239497,1.17458291){\makebox(0,0)[lt]{\lineheight{1.25}\smash{\begin{tabular}[t]{l}Narrowed\end{tabular}}}}%
    \put(0.35931759,1.17458291){\makebox(0,0)[lt]{\lineheight{1.25}\smash{\begin{tabular}[t]{l}chip\end{tabular}}}}%
    \put(0.45029347,1.17458291){\makebox(0,0)[lt]{\lineheight{1.25}\smash{\begin{tabular}[t]{l}mouth\end{tabular}}}}%
    \put(0.58759698,1.17458291){\makebox(0,0)[lt]{\lineheight{1.25}\smash{\begin{tabular}[t]{l}Extended\end{tabular}}}}%
    \put(0.77357688,1.17458291){\makebox(0,0)[lt]{\lineheight{1.25}\smash{\begin{tabular}[t]{l}chip\end{tabular}}}}%
    \put(0.86455176,1.17458291){\makebox(0,0)[lt]{\lineheight{1.25}\smash{\begin{tabular}[t]{l}mouth\end{tabular}}}}%
    \put(0.0953407,0.83990955){\rotatebox{90}{\makebox(0,0)[lt]{\lineheight{1.25}\smash{\begin{tabular}[t]{l}CAD model\end{tabular}}}}}%
    \put(0.06770251,0.4){\rotatebox{90}{\makebox(0,0)[lt]{\lineheight{1.25}\smash{\begin{tabular}[t]{l}Additively manuf.\end{tabular}}}}}%
    \put(0.12297889,0.41){\rotatebox{90}{\makebox(0,0)[lt]{\lineheight{1.25}\smash{\begin{tabular}[t]{l}drill head blanks \end{tabular}}}}}%
    \put(0.06770251,0.04){\rotatebox{90}{\makebox(0,0)[lt]{\lineheight{1.25}\smash{\begin{tabular}[t]{l}Drill heads after\end{tabular}}}}}%
    \put(0.12297889,0.05){\rotatebox{90}{\makebox(0,0)[lt]{\lineheight{1.25}\smash{\begin{tabular}[t]{l}post-processing\end{tabular}}}}}%
    \put(0,0){\includegraphics[width=\unitlength,page=2]{Fig7.pdf}}%
  \end{picture}%
\endgroup%

    \caption{CAD design, additive manufacturing, and post-processing of modified ejector drill heads.}
    \label{fig:manufactured drill heads}
\end{figure}

The modified drill heads are manufactured using laser powder bed fusion (LPBF) on a SLM 280 HL machine. Compared to the reference drill head, the modifications of the additively manufactured drill heads also include optimized coolant outlets that direct the MWF flow in a $20^\circ$ angle towards the cutting edges, guide pads, and chip mouths. Due to the high demands placed on the tolerances in the area of the cutting insert seats, the guide pad seats, and the connecting thread for the drill tube, it is necessary to machine these functional surfaces after the additive manufacturing process. Based on the CAD models of the optimized drill heads, adjustments were therefore made for the additive manufacturing process, which took into account a corresponding allowance of $\textrm{0.15}\,\textrm{mm}$ in these areas. The mounting thread for the drill tube as well as the threads for fastening the cutting inserts and guide pads are also only created during machining. Compared to the simulated geometry, minor adjustments were made to the edge of the chip mouth of modification II to ensure accessibility to the fastening screws of the outer cutting insert. In the additive process, the drill heads were manufactured in an upright position with the chip mouth facing upwards so that as few support structures as possible were required. The drill heads were made from the material X3NiCoMoTi18-9-5 with a layer thickness of $\textrm{s} = 0.03\,\textrm{mm}$. The build platform was preheated to a temperature of $\vartheta = 100\,^{\circ}\mathrm{C}$ to reduce thermal stresses and to ensure a good bond. The process pressure was $\textrm{p} = 12\,\textrm{mbar}$ with a gas flow velocity of $\textrm{v} = 20\,\textrm{m}\,\textrm{s}^{-1}$. Argon was used as process gas. The modifications were successfully implemented with the additive manufacturing process. In order to guarantee the application properties of the tool steel, the additive blanks underwent heat treatment before machining, in which they were solution-annealed at $\vartheta = 880\,^{\circ}\mathrm{C}$, quenched in water and then artificially aged at $\vartheta = 490\,^{\circ}\mathrm{C}$. Figure~\ref{fig:manufactured drill heads} shows the individual steps from the CAD model to the finished drill head with cutting edges and guide pads.

\subsection{Experimental Setup and Procedure}\label{subsec2}

The experimental analyses for ejector drilling with the reference drill head and the two modified variants were carried out on an Index G250 turn-mill center. On the machine with main and counter spindle, the workpiece was clamped in the main spindle and the ejector system in the counter spindle. The material used in the final tests was the heat-treated steel 42CrMo4+QT, which is frequently used for components subjected to higher loads. For a detailed analysis of the MWF supply, an external circuit was installed with a coolant pump controlled by a frequency converter, which was used to control the volume flow supplied during the process. A change in the pump setting was possible in $1\,\textrm{Hz}$ steps resulting in a reduction of the reference volume flow of $55\,\textrm{liter/min}$ by approx. $2.2 ... 2.4\,\textrm{liter/min}$ for each step. The volume flow itself was measured using appropriate sensors at the cooling lubricant outlet of the ejector system.

\begin{figure*}[ht!]
    \centering
    \def\svgwidth{\textwidth}

    \begingroup%
  \makeatletter%
  \providecommand\color[2][]{%
    \errmessage{(Inkscape) Color is used for the text in Inkscape, but the package 'color.sty' is not loaded}%
    \renewcommand\color[2][]{}%
  }%
  \providecommand\transparent[1]{%
    \errmessage{(Inkscape) Transparency is used (non-zero) for the text in Inkscape, but the package 'transparent.sty' is not loaded}%
    \renewcommand\transparent[1]{}%
  }%
  \providecommand\rotatebox[2]{#2}%
  \newcommand*\fsize{\dimexpr\f@size pt\relax}%
  \newcommand*\lineheight[1]{\fontsize{\fsize}{#1\fsize}\selectfont}%
  \ifx\svgwidth\undefined%
    \setlength{\unitlength}{1569bp}%
    \ifx\svgscale\undefined%
      \relax%
    \else%
      \setlength{\unitlength}{\unitlength * \real{\svgscale}}%
    \fi%
  \else%
    \setlength{\unitlength}{\svgwidth}%
  \fi%
  \global\let\svgwidth\undefined%
  \global\let\svgscale\undefined%
  \makeatother%
  \begin{picture}(1,0.54445507)%
    \lineheight{1}%
    \setlength\tabcolsep{0pt}%
    \put(0,0){\includegraphics[width=\unitlength,page=1]{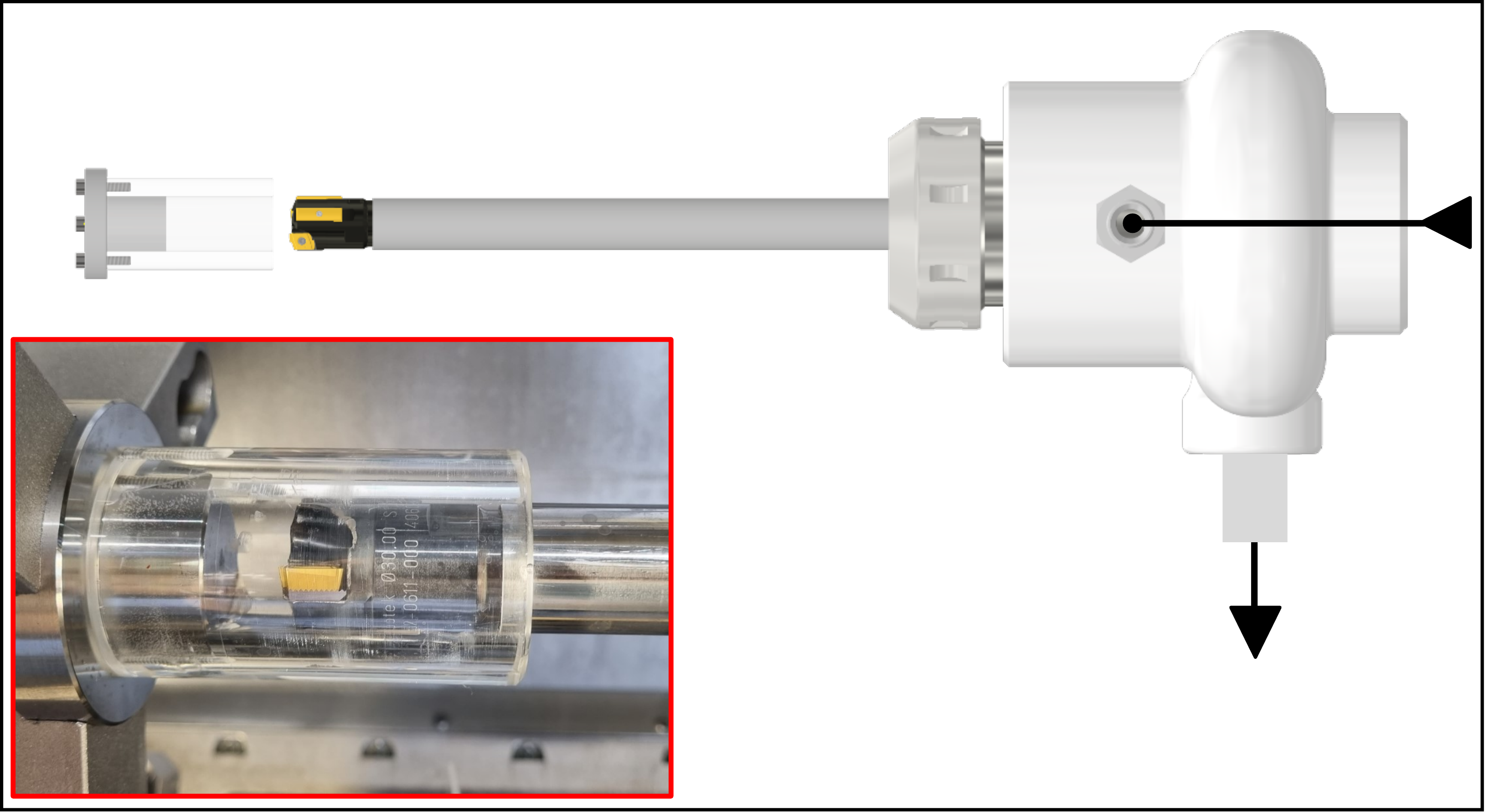}}%
    \put(0.64555529,0.50344566){\makebox(0,0)[lt]{\lineheight{1.25}\smash{\begin{tabular}[t]{l}Ejector system\end{tabular}}}}%
    \put(0.35,0.45){\makebox(0,0)[lt]{\lineheight{1.25}\smash{\begin{tabular}[t]{l}Drill tube\end{tabular}}}}%
    \put(0,0){\includegraphics[width=\unitlength,page=2]{Fig8.pdf}}%
    \put(0.27258397,0.50344566){\makebox(0,0)[lt]{\lineheight{1.25}\smash{\begin{tabular}[t]{l}Drill head\end{tabular}}}}%
    \put(0.01,0.50344566){\makebox(0,0)[lt]{\lineheight{1.25}\smash{\begin{tabular}[t]{l}Workpiece\end{tabular}}}}%
    \put(0.122,0.495){\makebox(0,0)[lt]{\lineheight{1.25}\smash{\begin{tabular}[t]{l}Polycarbonate\end{tabular}}}}%
    \put(0.15,0.47){\makebox(0,0)[lt]{\lineheight{1.25}\smash{\begin{tabular}[t]{l}sleeve \end{tabular}}}}%
    \put(0.47,0.495){\makebox(0,0)[lt]{\lineheight{1.25}\smash{\begin{tabular}[t]{l}Strain gauge \end{tabular}}}}%
    \put(0.5,0.47){\makebox(0,0)[lt]{\lineheight{1.25}\smash{\begin{tabular}[t]{l}array \end{tabular}}}}%
    \put(0,0){\includegraphics[width=\unitlength,page=3]{Fig8.pdf}}%
    \put(0.92720443,0.50344566){\makebox(0,0)[lt]{\lineheight{1.25}\smash{\begin{tabular}[t]{l}MWF\end{tabular}}}}%
    \put(0.93,0.48){\makebox(0,0)[lt]{\lineheight{1.25}\smash{\begin{tabular}[t]{l}inlet\end{tabular}}}}% 
    \put(0,0){\includegraphics[width=\unitlength,page=4]{Fig8.pdf}}%
    \put(0.90643005,0.15688735){\makebox(0,0)[lt]{\lineheight{1.25}\smash{\begin{tabular}[t]{l}MWF\end{tabular}}}}%
    \put(0.90643005,0.135){\makebox(0,0)[lt]{\lineheight{1.25}\smash{\begin{tabular}[t]{l}outlet\end{tabular}}}}%
    \put(0,0){\includegraphics[width=\unitlength,page=5]{Fig8.pdf}}%
  \end{picture}%
\endgroup%

    \caption{Ejector system with strain gauges for mechanical loads and polycarbonate sleeve for high-speed videos.}
    \label{fig:ejector system}
\end{figure*}

The aim of the analyses with the different drill heads was to determine the minimum volume flow at which a stable drilling process can be realized without chip blockage. A chip blockage results on the one hand in changes of the cooling lubricant flow at the drill head, and on the other hand the feed forces of the drilling process increase significantly. In order to define an appropriate termination criterion for the determination of the minimal necessary MWF flow, analyses of the chip removal at the drill head and the resulting increase in mechanical loads for a chip blockage were carried out for the reference drill head. In these preliminary tests, the material 18CrNiMo7-6 was used, which is prone to form long chips and thus favors the formation of chip blockages. Figure~\ref{fig:ejector system} shows an overview of the test setup. For the analyses of the chip removal with a Keyence VW 9000D high-speed camera, the bore hole wall for a workpiece sample with a diameter of \mbox{$\textrm{d} = 30\,\textrm{mm}$} was replaced by a transparent polycarbonate tube. On the one hand, this provides a drilling guide for the drilling process and, on the other hand, the polycarbonate tube allows optical access to the drill head during the drilling process. Since the MWF emulsion would block the optical access to the chip removal, it was replaced with water during these high-speed analyses. To measure the feed force and the drilling torque, strain gauges were used on the outer drill tube, which, coupled with a telemetry system, enabled wireless transmission of the measured values with a rotating tool.

The investigations to determine the minimal MWF flow rate were carried out with 42CrMo4+QT workpieces with a diameter of \mbox{$\textrm{d} = 50\,\textrm{mm}$}. To support the drill head at the beginning of the ejector drilling process, a guide bush was clamped in a special tool holder and placed in front of the sample with the turret of the turn-mill center.

\subsection{Chip Blockage Analysis}\label{subsec3}

Due to the large number of tests required for the reference tool and the modified drill heads to identify the minimum volume flow for a stable drilling process, it is not possible to identify chip blockages in each case using the relatively complex high-speed analyses of the process. It is therefore necessary to define appropriate critical values for the identification of chip blockages in the drilling process in advance of the tests via the process-parallel recording of torque or feed force. To specifically provoke a chip blockage for the analysis of the correlations, the material 18CrNiMo7-6 was machined with the reference drill head in preliminary tests with high-speed analyses. The tests were carried out at an increased cutting speed of $\textrm{v}_{c} = 80\,\textrm{m/min}$ and a low feed rate of $\textrm{f} = 0.1\,\textrm{mm}$ in order to force the formation of long chips. The MWF flow rate was selected on the basis of the parameter recommendations by the tool manufacturer for the process and amounted to \mbox{$\dot{\textrm{V}}= 55\,\textrm{liter/min}$}.

\begin{figure}[htb]
    \includegraphics[width=\linewidth,keepaspectratio]{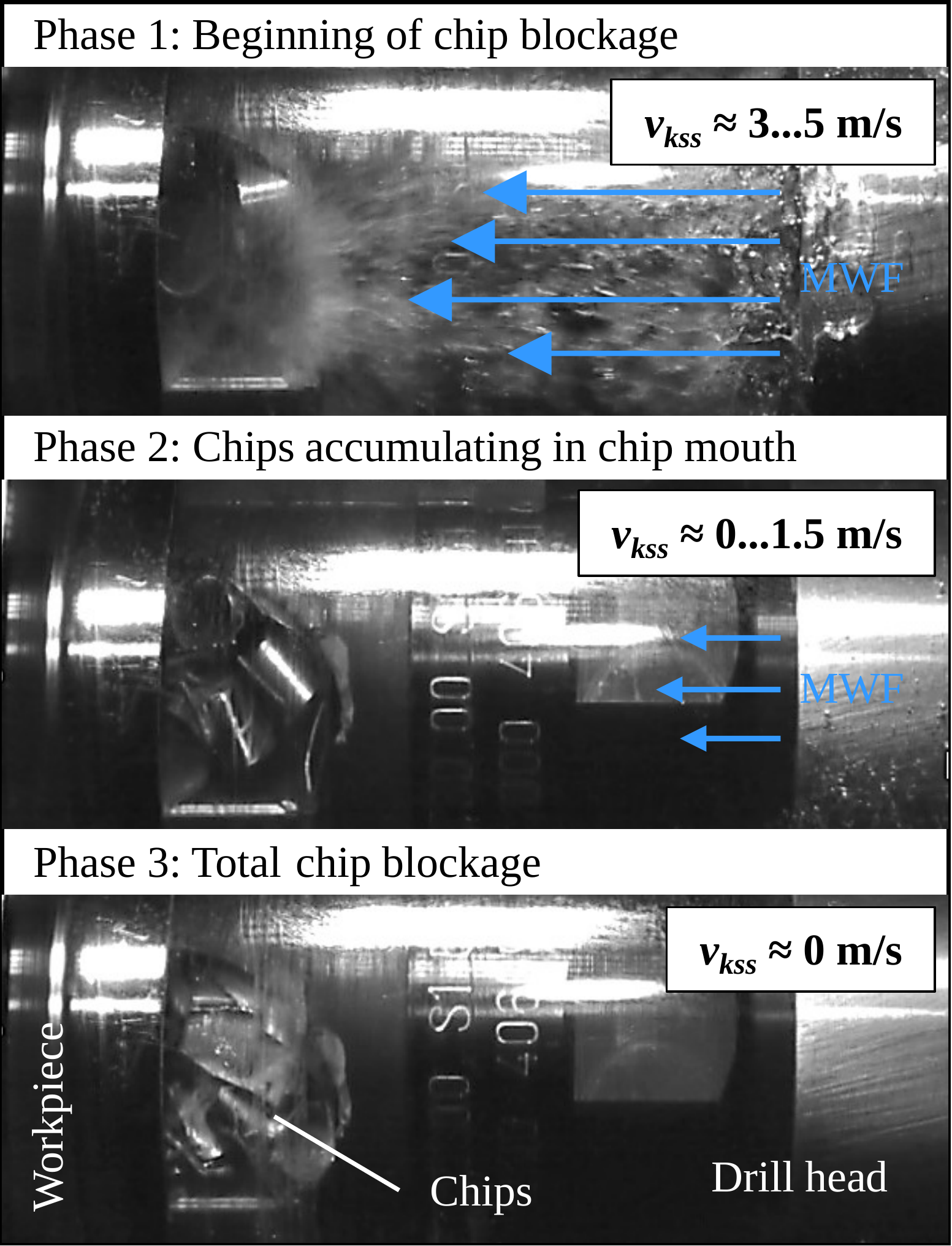}
    \caption{Stages of chip blockage.}
    \label{fig:chip blockage}
\end{figure}

Figure~\ref{fig:chip blockage} shows the formation of a chip blockage in still images of the high-speed video analyses. The formation of the chip blockage can be divided into three phases. In the first phase, a longer chip that forms on the outer cutting edge is not guided directly into the chip mouth by the MWF flow and is not discharged through the inner drill tube, but remains in the area of the chip mouth. At this point, the MWF flow velocity in the direction of the chip mouth is still relatively high with local fluid velocities of $\textrm{v} = 3...5\,\textrm{m}\,\textrm{s}^{-1}$. In the second phase, further chips accumulate on the first chip, which slow down the MWF flow and promote further jamming of the chips in the area of the chip mouth. In the third phase, a complete chip blockage occurs in the area of the outer chip mouth. The jammed chips block the MWF flow through the chip mouth and also prevent the continuous formation of further chips. The entire process from the first chip, which is not removed through the chip mouth, to the complete chip blockage is relatively short and, in the case shown, took place within just 7 tool rotations. With the help of the mechanical tool loads measured at the same time, it was also possible to prove the correlation between chip blockage and the rapid increase in feed force. With regard to the reliable identification of a complete chip blockage and the axial load on the machine spindles, a maximum feed force of $\textrm{F} = 6\,\textrm{kN}$ was defined as a termination criterion for the experiments to determine the minimum volume flow for the modified ejector drill heads. A typical feed force during usual operation would be around $\textrm{F} = 2.5\cdots\,4\,\textrm{kN}$~\cite{Gerken102024}.

The high-speed analyses show how important it is to remove chips quickly and smoothly with a targeted cooling lubricant flow toward the chip mouth in order to achieve a high level of process reliability in ejector drilling. They also confirm the flow analyses carried out with the SPH simulations in Chapter 3 with the selection of modification II and modification IV for a vortex-free MWF flow and the realization of the highest possible fluid velocities in the area of the chip mouth to optimize the ejector drilling process.

\subsection{Modified Ejector Drill Heads}\label{subsec4}

In order to investigate the general performance of the additively manufactured modified ejector drill heads, the tools were first tested with the parameters recommended by the tool manufacturer for the reference tool. For machining the material 42CrMo4+Qt, a cutting speed of \mbox{$\textrm{v}_{c} = 60 \,\textrm{m/min}$}, a feed rate of \mbox{$\textrm{f} = 0.2\,\textrm{mm}$}, and a MWF flow rate of \mbox{$\dot{\textrm{V}} = 55\,\textrm{liter/min}$} were used. The aim was to gain experience of the operational behavior, identify possible weak points and investigate wear of the drill heads over a drilling path of \mbox{$\textrm{l}_{f} = 25.000\,\textrm{mm}$}.

\begin{figure}[htb]
    \includegraphics[width=\linewidth,keepaspectratio]{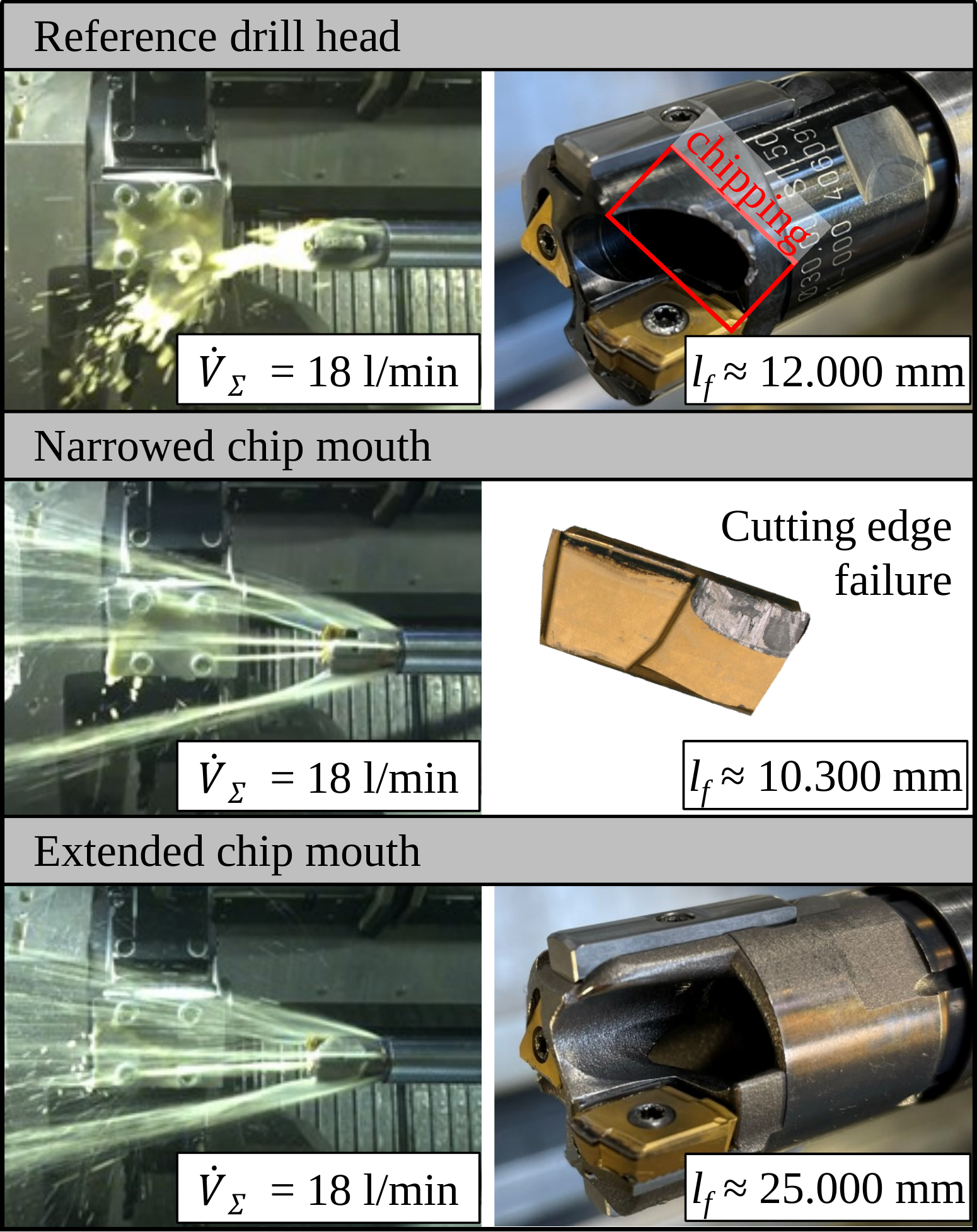}
    \caption{Comparison of MWF flow and tool life for the reference tool and the modified drill heads.}
    \label{fig:MWF flow}
\end{figure}

The positive influence of the modified MWF outlet holes with an angle of $20\,^{\circ}$ for the additive manufactured tools was already demonstrated during the initial setup of the tools where the effect of extremely low volume flows at the dill head were tested. The left-hand side of Figure~\ref{fig:MWF flow} shows the MWF flow for the reference tool and the two modified tools. It can be clearly seen that even for a comparatively low flow rate of  $\dot{\textrm{V}} = 18\,\textrm{liter/min}$, the modified tools produce a targeted jet of MWF in the direction of the tool tip and thus in the direction of the tool cutting edges, guide pads, bore hole bottom, and chip mouth. The direction of the flow is in very good agreement with the results of the detailed flow analysis using the possibilities of the SPH flow simulations when designing the modified ejector drill heads.

The drilling tests initially resulted in a stable process for all tools. In the case of the reference tool, minor chipping occurred at the edge of the chip mouth at approx. $\textrm{l}_{f} = 12.000\,\textrm{mm}$ drill path (see right-hand side Figure~\ref{fig:MWF flow}). The modified tool with a narrowed chip mouth repeatedly resulted in failure of the outer cutting edge, which could be attributed to chips jamming between the edge of the chip mouth and the cutting edge. The use of the modified tool with an extended chip mouth resulted in a stable drilling process with no visible signs of wear on the drill head. Since the tool with a reduced chip mouth resulted in chipping of the cutting edge and therefore a stable process could not be realized even with the standard parameters, the analyses to determine the minimum required volume flow were only carried out with the reference tool and the modified tool with extended chip mouth.

The minimum volume flow required for a stable drilling process for the two drill heads was determined taking into account the previously defined criterion of a maximum feed force of $\textrm{F} = 6\,\textrm{kN}$ for the identification of chip blockage in the process. Table~\ref{tab:MWF flow} shows the volume flows determined during the tests for the standard parameters of $\textrm{v}_{c} = 60\,\textrm{m/min}$ and \mbox{$\textrm{f} = 0.2\,\textrm{mm}$} as well as the results for an increased cutting speed of $\textrm{v}_{c} = 80\,\textrm{m/min}$. The tests show that with the modified tool for a cutting speed of $\textrm{v}_{c} = 60\,\textrm{m/min}$, the volume flow required for a stable process without chip blockage could be reduced by approx. $16\%$. For the increased cutting speed, the volume flow could be reduced from $\dot{\textrm{V}} = 51.5\,\textrm{liter/min}$ for the reference tool to $\dot{\textrm{V}} = 29.5\,\textrm{liter/min}$ for the modified tool. This corresponds to a reduction in the required volume flow of approx. $42\%$. The increase in the efficiency of the ejector drilling process results both from the adaptation of the coolant outlets and from the expansion of the chip mouth. The experimental investigations thus validate the findings from the SPH flow simulations that outlet holes at an angle of \mbox{$20^\circ$} lead to a directional supply of the fluid and that significantly higher local fluid velocities can be realized by extending the chip mouth, which favors the safe removal of the chips via the chip mouth into the inner drill tube.

\begin{table}[ht!]
\caption{Minimum MWF flow for a stable ejector drilling process without chip blockage.}
\label{tab:MWF flow}
\centering
\renewcommand{\arraystretch}{1.25}
\setlength{\tabcolsep}{6pt}  % <- Adjusts horizontal spacing
\begin{tabularx}{\columnwidth}{>{\hsize=1.05\hsize}p{1.5cm} 
                                  >{\centering\arraybackslash}X 
                                  >{\centering\arraybackslash}X}
\toprule
& $\mathbf{v_c = 60\,\mathrm{m/min}}$ & $\mathbf{v_c = 80\,\mathrm{m/min}}$ \\
\midrule
\wrap{\rule{0pt}{8pt}reference drill head}     & \mbox{$\dot{V}_\Sigma \approx 29.5\,\mathrm{liter/min}$} & \mbox{$\dot{V}_\Sigma \approx 51.5\,\mathrm{liter/min}$} \\
\cdashline{1-3}
\wrap{\rule{0pt}{12pt}extended chip mouth}       & \mbox{$\dot{V}_\Sigma \approx 24.7\,\mathrm{liter/min}$} & \mbox{$\dot{V}_\Sigma \approx 29.5\,\mathrm{liter/min}$} \\
\midrule
\wrap{\rule{0pt}{10pt}\textbf{MWF reduction}}    & \textbf{16.27\%} & \textbf{42.72\%} \\
%\wrap{\rule{0pt}{10pt}\textbf{MWF reduction}}    & \textbf{\colorbox{lime!80}{16.27\%}} & \textbf{\colorbox{lime!80}{42.72\%}} \\
\bottomrule
\end{tabularx}
\end{table}

\section{Conclusion}\label{sec5}
Ejector deep hole drilling offers in particular great potential to utilize the typical advantages of deep hole drilling processes on conventional machining centers for industrial applications in a cost-effective and resource-efficient manner. However, the current application in industrial use is associated with an inefficient use of resources due to the lack of process knowledge about the physical processes involved in ejector deep hole drilling, as the MWF volume flow is often set significantly higher than necessary.

In order to optimize the ejector deep hole drilling process, two modified drill head designs were numerically evaluated using the SPH method for their chip evacuation performance. They were additively manufactured and compared with the reference drill head in experimental investigations. The focus of the modifications on one hand was an extended chip mouth opening, which should ensure continuous efficient removal of chips and counteract the formation of chip blockages due to its faster flow rates. On the other hand, a narrowed chip mouth opening should contribute to decimation of the flow vortices at the outer cutting edge and thus ensure a targeted MWF supply into the cutting edge area, enhancing the chip evacuation from the cutting zone resulting from a streamlined flow.

Although the modification with a narrowed chip mouth resulted in less vortex formation in the area of the cutting edges, the volume flow was also reduced. In conjunction with the larger interfering contour for chip evacuation from the cutting edge, this resulted in local chip accumulation and chipping on the outer cutting edge. The modification with the enlarged chip mouth in combination with the tilting of the coolant outlet holes to $20^\circ$ in feed direction resulted in a nearly laminar flow pattern with improved supply to the cutting edges and guide pads as well as improved chip removal. Overall, the optimized tool was able to reduce the minimum volume flow required for a stable drilling process without chip blockage by approximately $43\%$. The results show very good agreement with the SPH simulations of the MWF flow and chip transport and form the basis for further optimization of the tools.

Although the optimization of the MWF outlet and the chip mouth shape on the drill head enabled lower volume flows, the analytic calculation~\cite{Astakhov1995} of the hydraulic efficiency range of the tool system with $\eta_e = 0.076 \ldots 0.107$ still show great potential for optimization. The modification of the ejector nozzles of the inner tube in particular offers the possibility for further optimizations of the tool system approaching the maximum hydraulic efficiency of $\eta_{e,\mathrm{max}} \approx 0.2$ without falling below the cavitation limit in the inner tube.

 Therefore, the focus for the continued developments is on the design of energy-efficient, modular ejector nozzle adapters that further reduce the MWF flow rate required to maintain the ejector effect and maximize the hydraulic efficiency $\eta_e$ through targeted adaptation of the ejector nozzle shape. In order to ensure maximum design freedom without the restrictions of conventional manufacturing processes and the rapid realization of different variants, the modular ejector nozzle adapters are also to be additively manufactured. In addition to additive manufacturing, laser material processing or micro-milling can also be used to realize complex designs. The modified ejector nozzles can be adapted to the original drill tube by means of a soldered or bonded connection and thus integrated into the standard tool system without major design changes. The previous results for the modification of the drill head and the planned modifications to the ejector nozzles contribute to the optimization of the whole ejector deep hole drilling system. Simulations of the process and experimental analyses work hand in hand to realize a more economical and resource-efficient use of ejector drilling in industrial applications.

\backmatter

% \bmhead{Acknowledgements}
% This research was supported by the Deutsche Forschungsgemeinschaft (DFG) under grant numbers 439917965 and 405605200.

\section*{Declarations}

\bmhead{Funding}
This research was supported by the Deutsche Forschungsgemeinschaft (DFG) under grant numbers 439917965 and 405605200.

\bmhead{Author contribution}
N.R.\, performed the numerical simulations, J.G.\, performed the experimental investigations, N.R.\,, S.M.\,, A.B.\,, and S.G.\, wrote the main manuscript text, P.E.\,, D.B.\,, S.M., and A.B.\, supervised the research, and corrected the manuscript. All authors discussed the findings and reviewed the manuscript.

\bmhead{Competing interests}
The authors declare that they have no conflict of interest.

\bmhead{Ethics approval}
Not applicable.

\bmhead{Consent for publication}
All authors approve the manuscript and give their consent for submission and publication.

\bmhead{Data availability}
The data that support the findings of this study are available from the corresponding author upon reasonable request.


\end{document}